\documentclass{JHEP3}
\usepackage{epsfig}
\usepackage{latexsym}  

\received{}
\revised{}
\accepted{}
\preprint{}

\newcommand{\D}{{\cal D}}
\newcommand{\PD}{\partial}
\newcommand{\TR}{{\mbox{Tr}}}
\newcommand{\sbra}[1]{\left(#1\right)}

\title{\large Gauge Problem of Monopole Dynamics in SU(2) Lattice Gauge Theory}
\author{Shoichi Ito, Tae Woong Park and Tsuneo Suzuki\\
Institute for Theoretical Physics, Kanazawa University,\\
Kanazawa 920-1192, Japan\\
E-mail: \email{shoichi@hep.s.kanazawa-u.ac.jp}\\
E-mail: \email{yasuo@hep.s.kanazawa-u.ac.jp}\\
E-mail: \email{suzuki@hep.s.kanazawa-u.ac.jp}}
\author{Shun-ichi Kitahara\\ Jumonji University,\\
Niiza, Saitama 352-8510, Japan\\
E-mail: \email{kitahara@jumonji-u.ac.jp}}

\abstract{
Gauge problem of monopole dynamics is studied in $SU(2)$ lattice gauge
theory.  We study first abelian and monopole contributions to the
static potential in four smooth gauges, i.e., Laplacian Abelian (LA),
Maximally Abelian Wilson Loop (MAWL) and L-type gauges in comparison
with Maximally Abelian (MA) gauge.  They all reproduce the string
tension in good agreement with the $SU(2)$ string tension.  MA gauge
is not the only choice of the good gauge which is suitable for the
color confinement mechanism.  Using an inverse Monte-Carlo method and
the blockspin transformation, we determine effective monopole actions
and the renormalization group (RG) flows of its coupling constants in
various abelian projection schemes.
Every RG flow looks to converge to a unique curve
which suggests gauge independence in the infrared region.
}


\begin{document}


\section{Introduction}
It is important to understand color confinement mechanism in Quantum
Chromodynamics (QCD).  Many numerical simulations have been done and
they support the dual superconductor scenario of the QCD vacuum as a
confinement mechanism~\cite{thooft76,Mandelstam76}.
Magnetic monopoles are induced by performing an abelian
projection~\cite{thooft}, i.e., a partial gauge-fixing that keeps
$U(1)\otimes U(1)$.  It is known that the string tension calculated
from the abelian and the monopole parts reproduces well the original
one when we perform an abelian projection in Maximally Abelian (MA)
gauge where link variables are abelianized as much as possible.  In
addition to the string tension, many low energy physical properties of
QCD are reproduced from the abelian and the monopole degrees of
freedom alone.  It is called as ``abelian and monopole dominance''.
These facts suggest that monopoles play an important role for the
confinement mechanism.  Actually, a low energy effective theory which
is described in terms of monopole currents has been derived by Shiba
and Suzuki~\cite{SS1} and an almost perfect monopole action showing
the scaling behavior is derived by Chernodub et al~\cite{Kato}.
Monopole condensation occurs due to energy-entropy balance~\cite{SS1}.
Abelian color-electric flux is squeezed into a string-like
shape~\cite{Bali,Koma} by the superconducting monopole current.  This
squeezed color flux causes a confinement potential between quarks.

We note that we have infinite degrees of freedom when we perform an
abelian projection.  That is to say, which gauge should be chosen?
Recently Laplacian Abelian (LA) gauge was proposed and it appears to
have similar good properties~\cite{Sijs1,Sijs2}. Actually MA and LA
gauges are very similar.  Are MA and LA gauges exceptional? If such is
the case, there must exist any reason to justify it, although it seems
very difficult to find such a reason.  Another interpretation is that
monopole dynamics does not depend on the choice of gauge in the
continuum limit, although it seems dependent of the gauge choice at
the present stage of lattice study.  In other words, MA gauge and LA
gauge are considered to have a wider window even at present to see the
continuum limit than other gauges have.

Our aim of this paper is to show first that MA gauge is not a special
choice of the good gauge for color confinement.  We restrict ourselves
to pure $SU(2)$ QCD for simplicity. Here we discuss two new gauges in
addition to LA gauge. They have a different continuum limit but they
all can reproduce well the $SU(2)$ string tension.  The second is to
derive an effective monopole action and to study the blockspin
transformation of the monopole currents in various abelian
projections.  If their renormalization group (RG) flows converge onto
the same line with a finite number of blockspin transformations, we
can expect gauge independence of monopole dynamics in the infrared
region.  The paper is organized as follows: In Section 2, we present
some theoretical and phenomenological arguments which support gauge
independence of abelian and monopole dominance.  In Section 3, we
describe gauge fixing procedures being used.  In Section 4, we show
that the $SU(2)$ string tension is well-reproduced from abelian or
monopole degrees of freedom alone in four different abelian projection
schemes.  In section 5, we present our results from RG flow study of
effective monopole actions in various abelian projections.  In Section
6, we summarize our conclusions.


\section{Theoretical and phenomenological background}
\subsection{Gauge fixings and abelian dominance}
It is known that the abelian Wilson loop reproduces well the $SU(2)$
string tension numerically, if MA gauge or LA gauge is
applied~\cite{Sijs2,SY}.  In the case of Polyakov gauge, the string
tension which is calculated from abelian Polyakov loop correlators is
exactly the same as that of $SU(2)$ ~\cite{Yasuta2}.  Shoji et
al. have developed a stochastic gauge fixing method which interpolates
between the MA gauge and no gauge fixing~\cite{Shoji2}.  They have
found that abelian dominance for the heavy quark potential is realized
even in a gauge which is far from the MA gauge.  In a finite
temperature system, abelian Polyakov loops in various gauges reproduce
the phase transition behavior of $SU(2)$ Polyakov
loop~\cite{Yotsuji}. See, Figure \ref{fig-aploop}.

Abelian dominance is shown also analytically.  Abelian Wilson loops
constructed without any gauge fixing give the same string tension as
that of $SU(2)$ Wilson loops in the strong coupling
expansion~\cite{SY}.  The same fact for any coupling region has been
proved by Ogilvie using the character expansion~\cite{Ogilvie}.  An
abelian Wilson loop operator is given by
\[ W_A\left[C\right]=\frac{1}{2}\TR\left[\prod_{s,\mu\in C}u_\mu(s)\right], \]
where $u_\mu$ is an abelian projected $U(1)$ link variable.  Since
$W_A$ is not a class function of $SU(2)$ group, only the $SU(2)$
invariant part extracted from $W_A$ is non-vanishing in the
expectation value.  This can be written as
\[ W_A^{inv}=\frac{1}{2}\int\D g\TR\left[
\prod_{s,\mu\in C}g(s)u_\mu(s)g^\dagger(s+\hat{\mu})\right]. \]
Using a character expansion, we get an expression for the expectation
value of the abelian Wilson loop in terms of $SU(2)$ Wilson loops:
\[ \langle W_A^{inv}\rangle=\left(\frac{2}{3}\right)^{P(C)}\langle 
W_{SU(2)}\rangle_{1/2}+\mbox{(half integer higher rep.).} \]
Since the lowest representation is dominant, we can show that the
$SU(2)$ string tension $\sigma_{SU(2)}$ can be reproduced perfectly
from the abelian string tension $\sigma_A$:
\[ \sigma_A=-\lim_{I,J\rightarrow\infty}\ln
\frac{\langle W_A(I+1,J+1)\rangle\langle W_A(I,J)\rangle}{\langle W_A(I+1,J)\rangle\langle W_A(I,J+1)\rangle}=\sigma_{SU(2)}. \]
Furthermore, Ogilvie has shown that similar arguments hold even
with a gauge fixing function
\[ S_{gf}=\lambda\sum\TR\left[U_\mu(s)\sigma_3U_\mu^\dagger(s)\sigma_3\right],\]
if the gauge parameter $\lambda$ is small enough.

\subsection{Monopole dominance}
There are numerical results supporting monopole dominance.  $SU(2)$
string tension is well reproduced only from the monopole part of
abelian Wilson loops in MA gauge~\cite{Stack,SS3} and LA
gauge~\cite{Sijs2}.  We note also that monopole Polyakov loops in
various gauges reproduce the phase transition behavior of $SU(2)$
Polyakov loop~\cite{Yotsuji}.  See, Figure \ref{fig-mploop}.

In addition to these numerical evidence, we can prove analytically
gauge independence of monopole dominance if abelian dominance is gauge
independent~\cite{lat99}.  If abelian dominance is gauge independent,
a common abelian effective action $S_{eff}$ written in terms of
abelian gauge field surely exists in any gauge and works well in the
infrared region as in MA gauge. Since $S_{eff}$ takes a form of
modified compact QED, an effective monopole action can be derived
analytically.  One can evaluate the contribution of monopoles to the
abelian Wilson loop using this effective monopole action.

In MA gauge, it is known numerically that an effective monopole action
composed of two-point self+Coulomb+nearest neighbor interactions is a
good approximation in the infrared region. The action can be
transformed exactly into a modified compact QED action in the generic
Villain form:
\[ Z=\int^\pi_{-\pi}\D\theta\sum_{n\in{\mathbb Z}}\exp\left[
-\frac{1}{4\pi^2}(d\theta+2\pi n,\Delta D(d\theta+2\pi n))+i(J,\theta)\right], \]
where $D\sim \beta\Delta^{-1}+\alpha+\gamma\Delta$.  An expectation
value of the abelian Wilson loop $W=e^{i(\theta,J)}$ can be estimated
using this action, where $J$ is the color electric current which takes
a value $\pm 1$ on a closed loop.  When we use the BKT
transformation~\cite{BKT1,BKT2}, we get an expectation value of the
abelian Wilson loop in terms of monopole currents $k$:
\[ \langle W\rangle=\frac{1}{Z}\sum_{k\in{\mathbb Z},dk=0}\exp\left[
-(k,Dk)-2\pi i(k,\delta\Delta^{-1}M)-\pi^2(J,(\Delta^2D)^{-1}J)\right], \]
where $M$ takes a value $\pm 1$ on a surface whose boundary is $J$
($J=\delta M$).  Electric-electric current ($J$-$J$) interactions are
of a modified Coulomb interaction and have no line singularity leading
to a linear potential. A linear potential of the abelian Wilson loop
originates from the second term of the monopole contribution.  Gauge
independence of monopole dominance is derived from that of abelian
dominance.  Gauge independence of an order parameter is also observed
in Ref.~\cite{DiGiacomo}.

\FIGURE{
\epsfig{scale=1.0,file=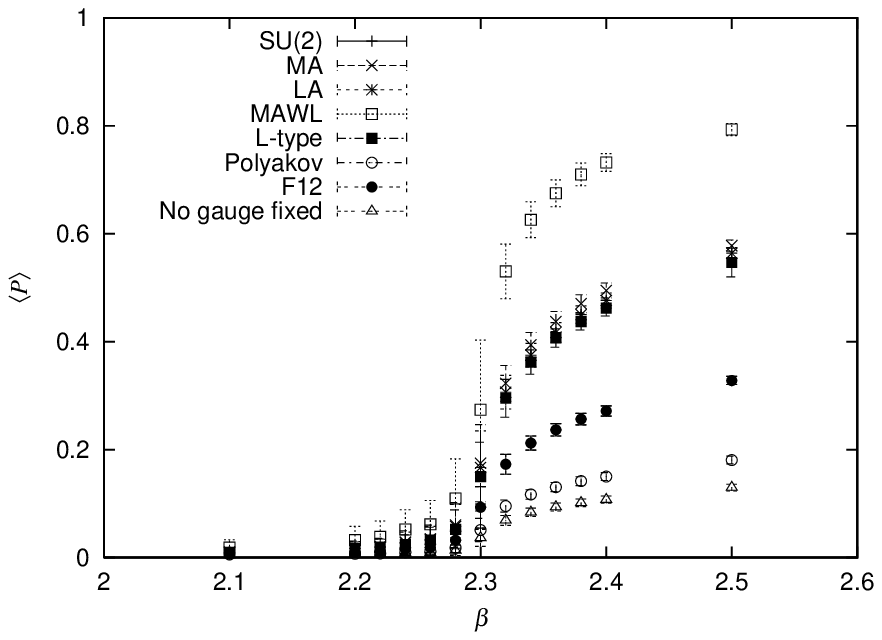}
\caption{$SU(2)$ Polyakov loop vs. abelian Polyakov loop in various gauges.
The behavior of the $SU(2)$ Polyakov loop is well reproduced
by the abelian Polyakov loop in various gauges.}
\label{fig-aploop}
}

\FIGURE{
\epsfig{scale=1.0,file=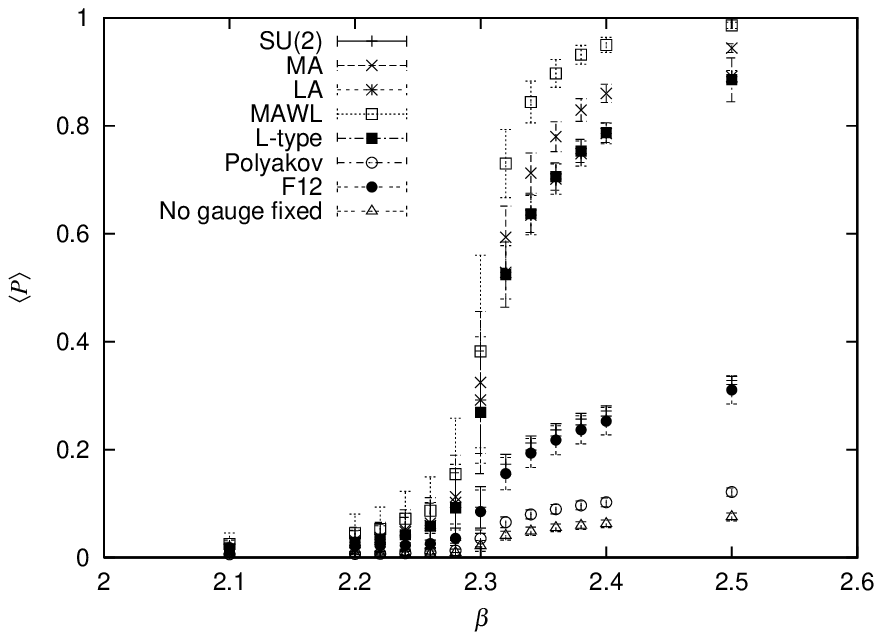}
\caption{$SU(2)$ Polyakov loop vs. monopole Polyakov loop in various gauges.
The behavior of the $SU(2)$ Polyakov loop is well reproduced
by the monopole Polyakov loop in various gauges.}
\label{fig-mploop}
}


\subsection{An objection to gauge independence}
As we have shown in previous subsections, there is encouraging
evidence which supports gauge independence of the confinement scenario
in terms of monopoles.  On the other hand, there is a strong objection
to the idea of gauge independence.

Consider a gauge called Polyakov gauge where Polyakov loop operators
are diagonalized in the continuum finite-temperature QCD.  It is
proved~\cite{Lenz,Chernodub} that the singularities of the gauge
fixing run only in the time-like direction.  This means that there are
only time-like monopoles in the system when the Polyakov gauge is
employed, if the degeneration points in abelian projection only
correspond to monopoles as 't Hooft argued.  Since such time-like
monopoles do not contribute to the physical string
tension~\cite{Matsubara}, monopole dominance is violated.

But numerically the above theoretical expectation seems to be
inconsistent with numerical data.  We show our preliminary result in
Figure \ref{fig-density2pol}.  The spatial and temporal monopole
densities are plotted in Figure \ref{fig-density2pol} as a function of
lattice spacing $a$ in the unit of physical string tension
$\sqrt{\sigma_p}$.  These densities are defined as
\[
\rho_s(\beta)=\frac{\frac{1}{3}\sum_{s}\sum_{i=1,2,3}|k_i(s)|}{(N_sa)^3N_4}
\hspace*{1em},\hspace*{1em}
\rho_t(\beta)=\frac{\sum_{s}|k_4(s)|}{(N_sa)^3N_4},
\]
respectively.
Figure \ref{fig-density2pol} shows that spatial (lattice) monopole
density may take non-zero value even in the $a\rightarrow 0$ limit.
This is not compatible with the theoretical expectation above.  In the
authors' opinion, the continuum limit of lattice monopoles must
contain extra ingredients different from the expected monopoles
corresponding to singularities of Polyakov loop operators.  We will
give a detailed analysis elsewhere.

\FIGURE{
\epsfig{scale=1.5,file=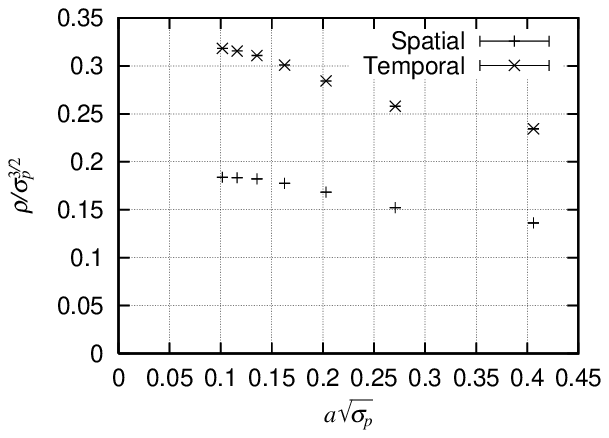}
\caption{Monopole density in Polyakov gauge versus lattice spacing}
\label{fig-density2pol}
}


\section{Various abelian projections on a lattice}
To check gauge (in)dependence of monopole dynamics, we study the
abelian projection in various gauges.
\begin{enumerate}
\item{MA gauge:\\ 
The most known is Maximally Abelian (MA) gauge. It is 
defined by maximizing the following quantity $R_{MA}$:
\begin{equation}
R_{MA}= \TR\sum_{s,\mu}U_\mu(s)\sigma_3U_\mu^\dagger(s)\sigma_3.
\label{eq:R1}
\end{equation}
This is achieved by 
diagonalizing an operator
\[ X_{MA}(s)=\sum_\mu\left(
U_\mu(s)\sigma_3U_\mu^\dagger(s)+
U_\mu^\dagger(s-\hat{\mu})\sigma_3U_\mu(s-\hat{\mu})
\right). \]
That is, 
\[ X_{MA}(s)\rightarrow X_{MA}'(s)
=V(s)X_{MA}(s)V^\dagger(s)
=\mbox{diag}\{\lambda_1,\lambda_2\},\]
where $V(s)$ is a gauge transformation matrix.
The diagonalization corresponds to the condition,
\begin{equation}
\sum_{\mu}\left(\PD_\mu\mp iA_\mu^3\right)A_\mu^\pm=0,
\end{equation}
in the continuum limit.
}
\item{LA gauge~\cite{Sijs1}:\\
First consider MA gauge again.
To maximize $R_{MA}$ in Eq.(\ref{eq:R1}) is to minimize the functional
\begin{eqnarray}
S_{MA}&=&\sum_{s,\mu}\left\{1-
\frac{1}{2}\TR[\Phi(s)U_\mu(s)\Phi(s+\hat{\mu})
U_\mu^\dagger(s)]\right\} \nonumber\\
&=&\sum_{s,\mu}\left\{\phi^a(s)R_\mu^{ab}(s)\phi^b(s+\hat{\mu})\right\},
\label{eq-lag-13}
\end{eqnarray}
where $R_\mu$ is a gauge field in the adjoint representation,
\[R_\mu^{ab}(s)=\frac{1}{2}\TR\left(
\sigma_aU_\mu(s)\sigma_bU_\mu^\dagger(s)\right).\]
$\Phi$ is parameterized by a spin variable $\phi$ which
satisfies a local constraint
\begin{equation}
\Phi(s)=V^\dagger(s)\sigma_3V(s)=\sum_{a=1}^3\phi^a(s)\sigma_a
\hspace*{1em},\hspace*{1em}
\sum_{a=1}^3\left(\phi^a(s)\right)^2=1.
\label{eq-lag-104}
\end{equation}
Because of the local constraint from the normalization,
it is very difficult to find a set of $\phi$ which
realizes the absolute minimum of Eq.(\ref{eq-lag-13}).

The key idea of the LA gauge fixing is to relax this constraint:
\[ \sum_{a=1}^3\left(\phi^a(s)\right)^2=1
\hspace*{1em}\longrightarrow\hspace*{1em}
\sum_s\sum_{a=1}^3\left(\phi^a(s)\right)^2=1 \]
The functional to minimize becomes
\begin{equation}
S_{LA}=\frac{1}{2}\sum_{x,a}\sum_{y,b}\phi^a(x)(-\Box_{xy}^{ab})\phi^b(y),
\label{eq-lag-108}
\end{equation}
where
\begin{equation}
-\Box_{xy}^{ab}=\sum_\mu\left(
2\delta_{xy}\delta^{ab}
-R_\mu^{ab}(x)\delta_{y,x+\hat{\mu}}
-R_\mu^{ba}(x-\hat{\mu})\delta_{y,x-\hat{\mu}}\right).
\label{eq-lag-109}
\end{equation}
Minimizing Eq.(\ref{eq-lag-108}) amounts to finding the eigenvector
belonging to the lowest eigenvalue of the covariant laplacian
operator.  This eigenvalue problem can be solved numerically (we used
an implicitly restarted Arnoldi method.  For example, see
Ref.~\cite{Arpack}).  The gauge transformation matrix $V(s)$ is
defined by
\begin{equation}
V^\dagger(s)\sigma_3V(s)=\sum_{a=1}^3\hat{\phi}^a(s)\sigma_a,
\end{equation}
where
\begin{equation}
\phi^a(s)=\rho(s)\hat{\phi}^a(s)
\hspace*{1em},\hspace*{1em}
\rho^2(s)=\sum_{a=1}^3\left(\phi^a(s)\right)^2.
\end{equation}
In the continuum limit, LA gauge corresponds to the gauge condition,
\begin{equation}
\sum_{\mu}\left(\PD_\mu\mp iA_\mu^3\right)\left(\rho^2A_\mu^\pm\right)=0.
\end{equation}
}
\item{MAWL gauge~\cite{MAWL}:\\
Maximally Abelian Wilson Loop (MAWL) gauge is a gauge which maximizes 
a Wilson loop operator written in terms of abelian link variables:
\begin{equation}
W_A=\cos\theta_{\mu\nu}(s),
\end{equation}
where
$\theta_{\mu\nu}(s)=\theta_{\mu}(s)+\theta_{\nu}(s+\hat{\mu})-\theta_{\mu}(s+\hat{\nu})-\theta_{\nu}(s)$.
It is achieved by diagonalizing the following operator
\begin{eqnarray*}
X_{MAWL}(s)&=&\sum_{\mu\ne\nu}\left[
\frac{\sin\theta_{\mu\nu}(s)-\sin\theta_{\mu\nu}(s-\hat{\nu})}{U_0^2(s,\mu)+U_3^2(s,\mu)}(U(s,\mu)\sigma_3U^\dagger(s,\mu))\right.\\
&&+\left.\strut\frac{\sin\theta_{\mu\nu}(s-\hat{\mu}-\hat{\nu})-\sin\theta_{\mu\nu}(s-\hat{\mu})}{U_0^2(s-\hat{\mu},\mu)+U_3^2(s-\hat{\mu},\mu)}(U^\dagger(s-\hat{\mu},\mu)\sigma_3U(s-\hat{\mu},\mu))
\right].
\label{eq-mawl-x}
\end{eqnarray*}

In the continuum limit, we get the following gauge condition:
\begin{equation}
\sum_{\mu\ne\nu}\partial_\nu f_{\mu\nu}A^{\pm}_{\mu}=0.
\end{equation}
}
\item{L-type gauge:\\
There are infinitely many gauges similar to MA gauge. Here we show one
simplest extension called L-type gauge. It is defined by maximizing
\begin{equation}
R_L= \TR\sum_{s,\mu\ne\nu}
U_\mu(s)\sigma_3U_\nu(s+\hat{\mu})\sigma_3
U_\nu^\dagger(s+\hat{\mu})\sigma_3U_\nu^\dagger(s)\sigma_3.
\end{equation}
This is given by diagonalizing 
\begin{eqnarray*}
X_L(s)&=&\sum_{\mu\ne\nu}\left[
U_{\mu}(s)\sigma_{3}U_{\mu}^{\dagger}(s)\sigma_{3}
U_{\nu}(s)\sigma_{3}U_{\nu}^{\dagger}(s)\right.\\
&&\left.+U_{\mu}^{\dagger}(s-\hat{\mu})\sigma_{3}U_{\nu}(s-\hat{\mu})\sigma_{3}
U_{\nu}^{\dagger}(s-\hat{\mu})\sigma_{3}U_{\mu}(s-\hat{\mu})
\right].
\end{eqnarray*}
A schematic representation of $R_L$
is shown in Figure \ref{fig-l_sch}.

In the continuum limit, we get the following gauge condition:
\begin{equation}
\sum_{\mu\ne\nu}\{(\partial_{\mu}\pm iagA^3_{\mu})
+(\partial_{\nu}\pm iagA^3_{\nu})\}(A^{\mp}_{\mu}+A^{\mp}_{\nu})=0.
\end{equation}
}
\item{ 
There are various gauges called unitary gauge. 
Polyakov gauge and F${}_{12}$ gauge are defined with
the following operators which are diagonalized:
\begin{eqnarray}
X_{Pol}(s)&=&\prod_{i=1}^{N_4}U_4(s+(i-1)\hat{4}),\label{eq-xpol}\\
X_{F_{12}}(s)&=&U_1(s)U_2(s+\hat{1})U_1^\dagger(s+\hat{2})U_2^\dagger(s),
\end{eqnarray}
respectively.

In the continuum, the Polyakov gauge is reduced to 
\begin{equation}
A^{\pm}_0(x)=0,
\end{equation}
whereas F${}_{12}$ gauge gives 
\begin{equation}
F_{12}^{\pm}(x)=0.
\end{equation}
}
\item{
We also consider simple abelian components extracted without
gauge-fixing, where exact abelian dominance is proved
analytically~\cite{Ogilvie}.}
\end{enumerate}

\FIGURE{
\epsfig{scale=0.8,file=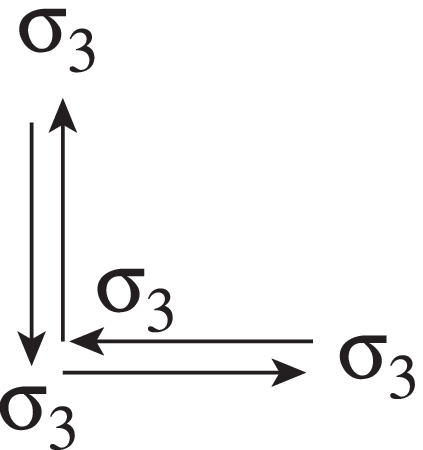}
\caption{Schematic representation of L-type gauge}
\label{fig-l_sch}
}


\section{String Tension}
As a first step, we measure abelian and monopole contributions to the
string tension in various abelian projections.  We used 100
configurations of $32^3\times 16$ lattice for the measurement.  In
this case, the critical point lies near $\beta_c\sim 2.7$.  We set the
gauge coupling $\beta$ to 2.5, so that the system is in the
confinement phase.  To reduce the statistical errors efficiently, we
adopted the hypercubic blocking~\cite{HYP} to the original
configurations.

The value of Polyakov loop correlators corresponds to the static
potential between one pair of quark and anti-quark;
\begin{equation}
\langle \TR P(0)\TR P^\dagger(R)\rangle =e^{-V(R)/T},
\label{eq-ppform}
\end{equation}
where $P(R)$ is the Polyakov loop operator Eq.(\ref{eq-xpol}).  $V(R)$
gives the inter-quark potential
\begin{equation}
V(R)=\sigma R-\frac{\alpha}{R}+c
\end{equation}
and $T=1/(N_4a)$ is the temperature of the system.

The abelian Polyakov loop operator is written as
\begin{equation}
P_a=\exp[i\sum_{i=0}^{N_4-1}\theta_4(\vec{s}+i\hat{4})].
\label{eq-ap-def}
\end{equation}
Eq.(\ref{eq-ap-def}) can be decomposed to photon and monopole
parts~\cite{Yasuta2} as follows:
\begin{eqnarray*}
P_a&=&P_p\cdot P_m,\\
P_p&=&\exp[-i\sum_{i=0}^{N_4-1}\sum_{s'}
D(\vec{s}+i\hat{4}-s')\PD_\nu'\bar{\Theta}_{\nu 4}(s')],\\
P_m&=&\exp[-2\pi i\sum_{i=0}^{N_4-1}\sum_{s'}
D(\vec{s}+i\hat{4}-s')\PD_\nu'n_{\nu 4}(s')],
\end{eqnarray*}
where we use an identity
\[ \theta_4(s)=-\sum_{s'}D(s-s')\left[\PD_\nu'\Theta_{\nu 4}(s')+\PD_4(\PD_\nu'\theta_\nu(s'))\right]. \]
The abelian field strength tensor
\[ \Theta_{\mu\nu}(s)\equiv\theta_\mu(s)+\theta_\nu(s+\hat{\mu})
-\theta_\mu(s+\hat{\nu})-\theta_\nu(s)\hspace*{1em},\hspace*{1em}
(-4\pi\le\theta_{\mu\nu}(s)<4\pi) \]
can be decomposed into two parts:
\[ \Theta_{\mu\nu}(s)\equiv\bar{\Theta}_{\mu\nu}(s)+2\pi n_{\mu\nu}(s)
\hspace*{1em},\hspace*{1em}(-\pi\le\bar{\Theta}_{\mu\nu}(s)<\pi). \]
Here, $\bar{\Theta}_{\mu\nu}(s)$ is interpreted as the
electro-magnetic flux through the plaquette and the integer valued
$n_{\mu\nu}(s)$ corresponds to the number of Dirac string piercing the
plaquette.
$D(s-s')$ is the Coulomb propagator on a lattice.

Figures \ref{fig-pc3MAGh250}, \ref{fig-pc3LAGh250},
\ref{fig-pc3MAWh250} and \ref{fig-pc3LLLh250}
show the values of $SU(2)$, abelian and monopole Polyakov loop
correlators in MA, LA, MAWL and L-type gauges, respectively.  The
values of abelian and monopole Polyakov loop correlators in each gauge
almost degenerate.  The string tension $\sigma$ can be extracted from
these values by fitting them to Eq.(\ref{eq-ppform}).  Fitted lines
are also plotted in the same figure.  In the case of MA gauge, fitted
values are consistent with the results by Bali et al~\cite{Bali2}.  In
the case of other gauges like a unitary gauge, one can not extract the
string tension clearly from the abelian and the monopole Polyakov loop
correlators due to large statistical errors.

\DOUBLEFIGURE[b]
{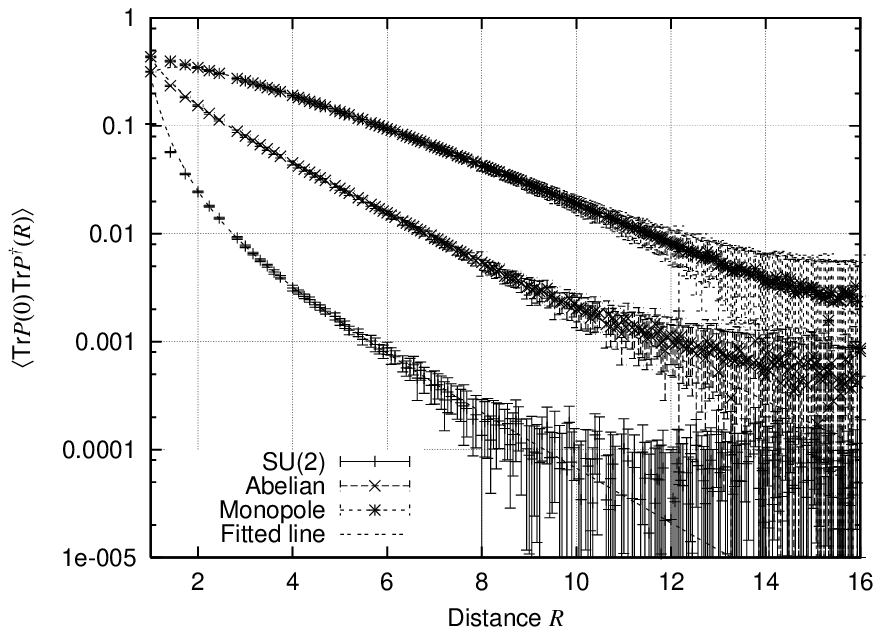,scale=0.7}
{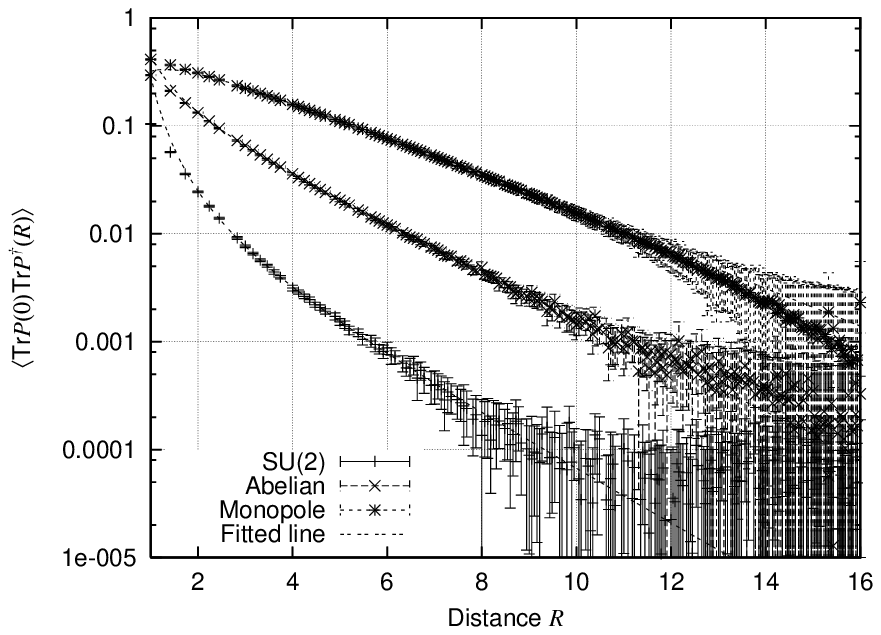,scale=0.7}
{{Abelian and monopole Polyakov loop correlator in MA gauge}\label{fig-pc3MAGh250}}
{{Abelian and monopole Polyakov loop correlator in LA gauge}\label{fig-pc3LAGh250}}

\DOUBLEFIGURE[b]
{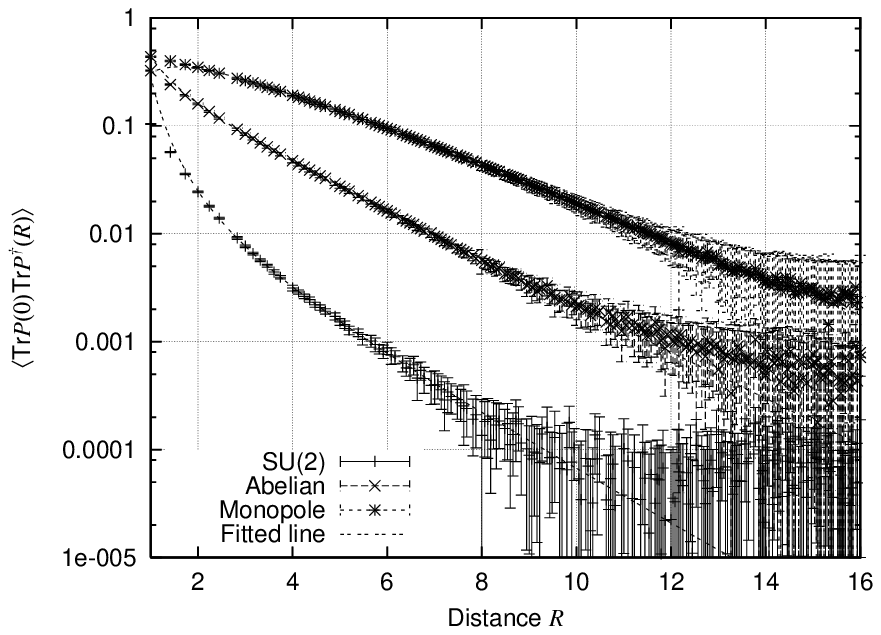,scale=0.7}
{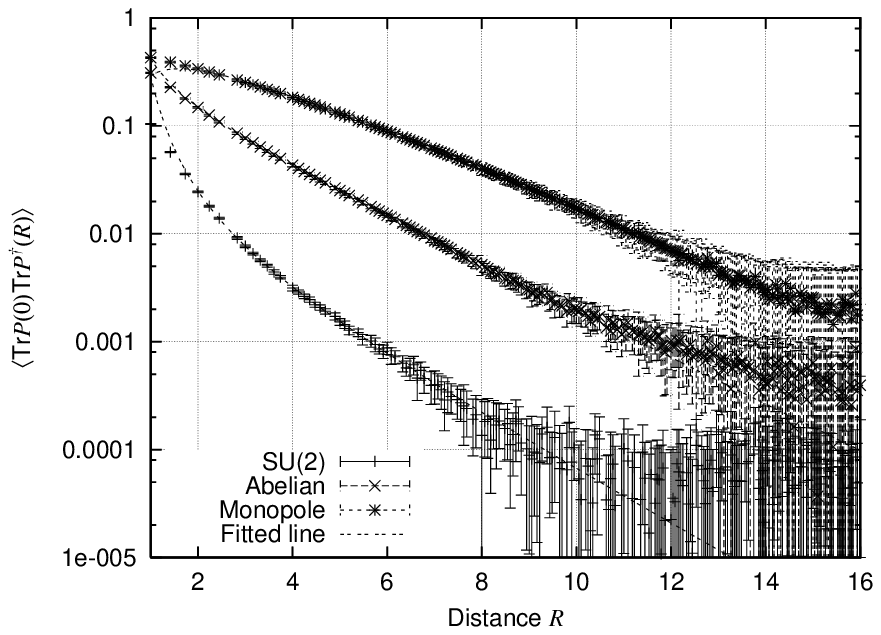,scale=0.7}
{{Abelian and monopole Polyakov loop correlator in MAWL gauge}\label{fig-pc3MAWh250}}
{{Abelian and monopole Polyakov loop correlator in L-type gauge}\label{fig-pc3LLLh250}}

Explicit values of the fitted string tension are shown in Table
\ref{tbl-fitted-st}.  They almost agree each other, although these
four gauges have different gauge fixing condition in the continuum
limit.

\TABLE{
\begin{tabular}{c|c|c|c|c}
 & MA & LA & MAWL & L-type\\
\hline
Abelian & 0.03054(45) & 0.03011(34) & 0.03051(45) & 0.03065(43) \\
\hline
Monopole & 0.02545(31) & 0.02536(28) & 0.02546(31) & 0.02624(34) \\
\end{tabular}
\caption{Fitted string tensions ($32^3\times 16$ lattice, $\beta=2.5$)
$\sigma_{SU(2)}=0.03446(105)$.}
\label{tbl-fitted-st}
}


\section{RG flows of the effective action in various abelian projections}
To clarify what is happening in the monopole dynamics, we study the
effective monopole actions in various gauges in this section.

\subsection{Simulation method}
Our method to derive an effective monopole action is the following.
We generate $SU(2)$ gauge fields $\{U_\mu(s)\}$ using the standard
$SU(2)$ Wilson action.  We consider $48^4$ hypercubic lattice for
$\beta$ from 2.1 to 2.5.  We took 50 independent configurations after
10,000 thermalization sweeps.  Then, we perform the abelian projection
in six different gauge fixings to extract abelian gauge fields
$\{u_\mu(s)\}$ from $SU(2)$ gauge fields.

One can define magnetic monopole currents from abelian field strength
tensor along the way of DeGrand and Toussaint~\cite{DGT}.
We can define the monopole current $k_\mu(s)$ as
\begin{equation}
k_\mu(s)=\frac{1}{2}\epsilon_{\mu\nu\rho\sigma}\PD_\nu n_{\rho\sigma}(s+\hat{\mu}).
\label{eq-kdef}
\end{equation}
By definition, it satisfies a current conservation law
\[ \PD_\mu'k_\mu(s)=0, \]
where $\PD_\mu$ and $\PD'_\mu$ denote the forward
and the backward differences in $\mu$-direction respectively.

We want to get an effective monopole action $S[k]$ on the dual lattice,
integrating out all degrees of freedom except for monopoles:
\begin{eqnarray*}
Z&=&\int\D Ue^{-S[U]}\delta(X^\pm)\Delta_F(U)\\
&=&\int\D u\left[\int\D Ce^{-S[U]}\delta(X^\pm)\Delta(U)\right]\\
&=&\int\D ue^{-S_{eff}[u]}\\
&=&\left(\prod\sum\right)\int\D ue^{-S_{eff}[u]}\delta(k,f(u))\\
&=&\prod_{s,\mu}\sum_{k_\mu(s)=-\infty}^{\infty}
\left(\prod_{m,\nu}\delta_{\PD_{\nu}'k_\nu(m),0}
\right)e^{-S[k]},
\end{eqnarray*}
where $U_\mu=C_\mu u_\mu$ and $X^\pm$ is the off-diagonal element of
the matrix $X$ which is diagonalized in the procedure of abelian
projection.  $\Delta_F(U)$ is the Faddeev-Popov determinant and
$\delta(k,f(u))$ gives the definition of the monopole current $k$ as a
function of abelian gauge field $u$.

Above integrations are done numerically.  We create vacuum ensembles
of monopole currents using the Monte-Carlo method and the definition
of the monopole current Eq.(\ref{eq-kdef}).  Then, we construct the
effective monopole action from monopole vacua using the Swendsen's
inverse Monte-Carlo method which was developed originally by
Swendsen~\cite{SS2} and extended by Shiba and Suzuki~\cite{SS1}.

We consider a set of independent and local monopole interactions which
are summed up over the whole lattice.  We denote each interaction term
as $S_i[k]$.  Then the effective monopole action can be written as a
linear combination of these operators:
\begin{equation}
S[k]=\sum_i g_iS_i[k],
\end{equation}
where $g_i$ denotes the effective coupling constants.  Explicit forms
of the interaction terms are listed in Table \ref{tbl:appquad} and
\ref{tbl:higher}.  We determine the set of couplings $\{g_i\}$ from
the monopole current ensemble $\{k_\mu(s)\}$ with the aid of an
inverse Monte-Carlo method.  Practically, we have to restrict the
number of interaction terms.  The form of action adopted here is 27
quadratic interactions and 4-point and 6-point
interactions~\cite{Kato,NKMR}.

We perform a blockspin transformation in terms of the monopole
currents on the dual lattice to study the RG flow.
The $n$-step blocked current is defined by
\begin{equation}
K_\mu(s^{(n)})=\sum_{i,j,l=0}^{n-1}k_\mu(ns^{(n)}+(n-1)\hat{\mu}
+i\hat{\nu}+j\hat{\rho}+l\hat{\sigma}).
\end{equation}
The blocked lattice spacing $b$ is given as $b=na(\beta)$ and the
continuum limit is taken as the limit $n\rightarrow\infty$ for a fixed
physical scale $b$.  We determine the effective monopole action from
the blocked monopole current ensemble $\{K_\mu(s^{(n)})\}$.  Then one
can obtain the RG flow in the 29-dimensional coupling constant space.

\TABLE{
\begin{tabular}{cllcll}
Coupling & Distance & \multicolumn{1}{c}{Type} &
Coupling & Distance & \multicolumn{1}{c}{Type} \\ 
\hline
$g_1$    & (0,0,0,0) & $k_\mu(s)$ &
$g_{15}$ & (2,1,1,0) & $k_\mu(s+2\hat{\mu}+\hat{\nu}+\hat{\rho})$ \\
$g_2$    & (1,0,0,0) & $k_\mu(s+\hat{\mu})$ &
$g_{16}$ & (1,2,1,0) & $k_\mu(s+\hat{\mu}+2\hat{\nu}+\hat{\rho})$ \\
$g_3$    & (0,1,0,0) & $k_\mu(s+\hat{\nu})$ &
$g_{17}$ & (0,2,1,1) & $k_\mu(s+2\hat{\nu}+\hat{\rho}+\hat{\sigma})$ \\
$g_4$    & (1,1,0,0) & $k_\mu(s+\hat{\mu}+\hat{\nu})$ &
$g_{18}$ & (2,1,1,1) & $k_\mu(s+2\hat{\mu}+\hat{\nu}+\hat{\rho}+\hat{\sigma})$ \\
$g_5$    & (0,1,1,0) & $k_\mu(s+\hat{\nu}+\hat{\rho})$ &
$g_{19}$ & (1,2,1,1) & $k_\mu(s+\hat{\mu}+2\hat{\nu}+\hat{\rho}+\hat{\sigma})$ \\
$g_6$    & (2,0,0,0) & $k_\mu(s+2\hat{\mu})$ &
$g_{20}$ & (2,2,0,0) & $k_\mu(s+2\hat{\mu}+2\hat{\nu})$ \\
$g_7$    & (0,2,0,0) & $k_\mu(s+2\hat{\nu})$ &
$g_{21}$ & (0,2,2,0) & $k_\mu(s+2\hat{\nu}+2\hat{\rho})$ \\
$g_8$    & (1,1,1,1) & $k_\mu(s+\hat{\mu}+\hat{\nu}+\hat{\rho}+\hat{\sigma})$ &
$g_{22}$ & (3,0,0,0) & $k_\mu(s+3\hat{\mu})$ \\
$g_9$    & (1,1,1,0) & $k_\mu(s+\hat{\mu}+\hat{\nu}+\hat{\rho})$ &
$g_{23}$ & (0,3,0,0) & $k_\mu(s+3\hat{\nu})$ \\
$g_{10}$ & (0,1,1,1) & $k_\mu(s+\hat{\nu}+\hat{\rho}+\hat{\sigma})$ &
$g_{24}$ & (2,2,1,0) & $k_\mu(s+2\hat{\mu}+2\hat{\nu}+\hat{\rho})$ \\
$g_{11}$ & (2,1,0,0) & $k_\mu(s+2\hat{\mu}+\hat{\nu})$ &
$g_{25}$ & (1,2,2,0) & $k_\mu(s+\hat{\mu}+2\hat{\nu}+2\hat{\rho})$ \\ 
$g_{12}$ & (1,2,0,0) & $k_\mu(s+\hat{\mu}+2\hat{\nu})$ &
$g_{26}$ & (0,2,2,1) & $k_\mu(s+2\hat{\nu}+2\hat{\rho}+\hat{\sigma})$ \\
$g_{13}$ & (0,2,1,0) & $k_\mu(s+2\hat{\nu}+\hat{\rho})$ &
$g_{27}$ & (2,2,1,0) & $k_\rho(s+2\hat{\mu}+2\hat{\nu}+\hat{\rho})$ \\
$g_{14}$ & (2,1,0,0) & $k_\nu(s+2\hat{\mu}+\hat{\nu})$ &
         &           & \\
\end{tabular}
\caption{The quadratic interactions used for the modified Swendsen's method.}
\label{tbl:appquad} 
}

\TABLE{
\begin{tabular}{cc}
Coupling & Type\\ 
\hline
4-point $g_{28}$ &
$\quad \quad \sum_{s}\sbra{\sum_{\mu=-4}^4 k_\mu^2(s)}^2$ \\
6-point $g_{29}$ &
$\quad \quad \sum_{s}\sbra{\sum_{\mu=-4}^4 k_\mu^2(s)}^3$ \\
\end{tabular}
\caption{The higher order interactions used for
the modified Swendsen's method.}
\label{tbl:higher} 
}

\subsection{Numerical Results}
The effective monopole action is determined successfully.  All
coupling constants which are contained in the effective monopole
action are obtained with relatively small errors.  We use the
jackknife method for the error estimation.  These effective monopole
actions except in MA gauge are determined for the first time in this
paper.  Moreover, these effective monopole actions are determined from
the blocked monopole configurations, too.  The results are summarized
as follows:
\begin{enumerate}
\item{
Only the quadratic interaction subspace seems sufficient in the
coupling space for the low-energy region of QCD.  Figure
\ref{fig-g28_vs_b} and \ref{fig-g29_vs_b} show coupling constants
\footnote{Effective coupling constants for the blocking factor $n=1$
are omitted in Figures \ref{fig-g28_vs_b},
\ref{fig-g29_vs_b} and \ref{fig-g1vsb1}-\ref{fig-flow4}.}
for 4-point and 6-point interaction terms versus physical scale $b$.
In the case of MA, LA, MAWL and L-type gauges, these coupling
constants take relatively larger absolute values for small $b$ region.
They become negligibly small for large $b$ region.  In the case of
Polyakov, F${}_{12}$, no gauge fixings, coupling constants for 4-point
and 6-point interaction terms take the values very close to zero in
the whole region of $b$.
}
\item{
Typical case of the coupling constants for quadratic interaction terms
versus squared distances in the lattice unit are shown in Figure
\ref{fig-r-dep}.  We see that coupling constants for the self
interaction term $g_1$ and the nearest-neighbor interactions $g_2$ and
$g_3$ are dominant, and $g_2\simeq g_3$.  Other couplings decrease
exponentially as distance between the two monopole currents grows.
Such a behavior does not depend on a gauge coupling constant $\beta$.
Therefore, we concentrate our analysis on the coupling constants of
quadratic interaction terms, especially $g_1$ and $g_2$.

\DOUBLEFIGURE[b]
{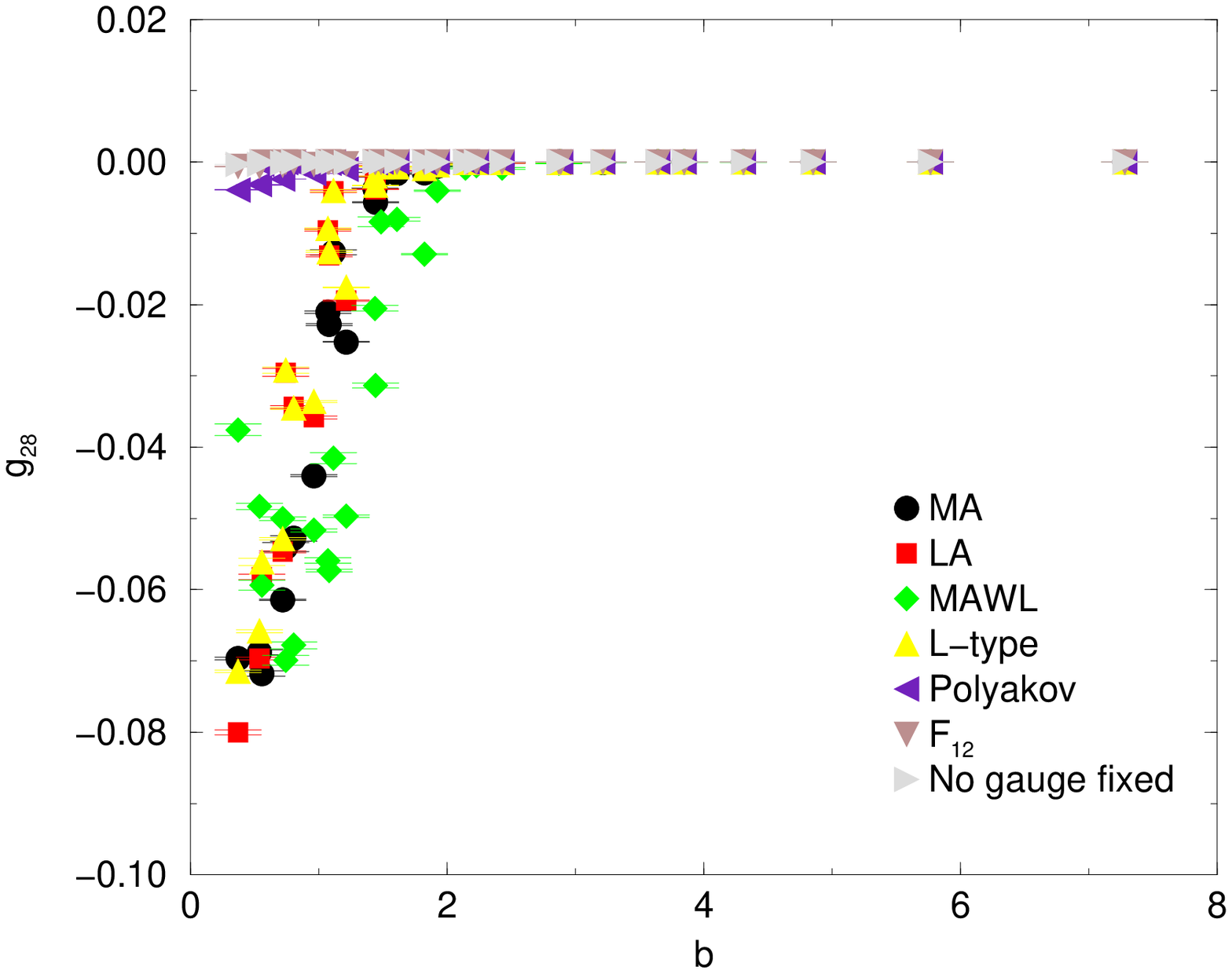,scale=0.41}
{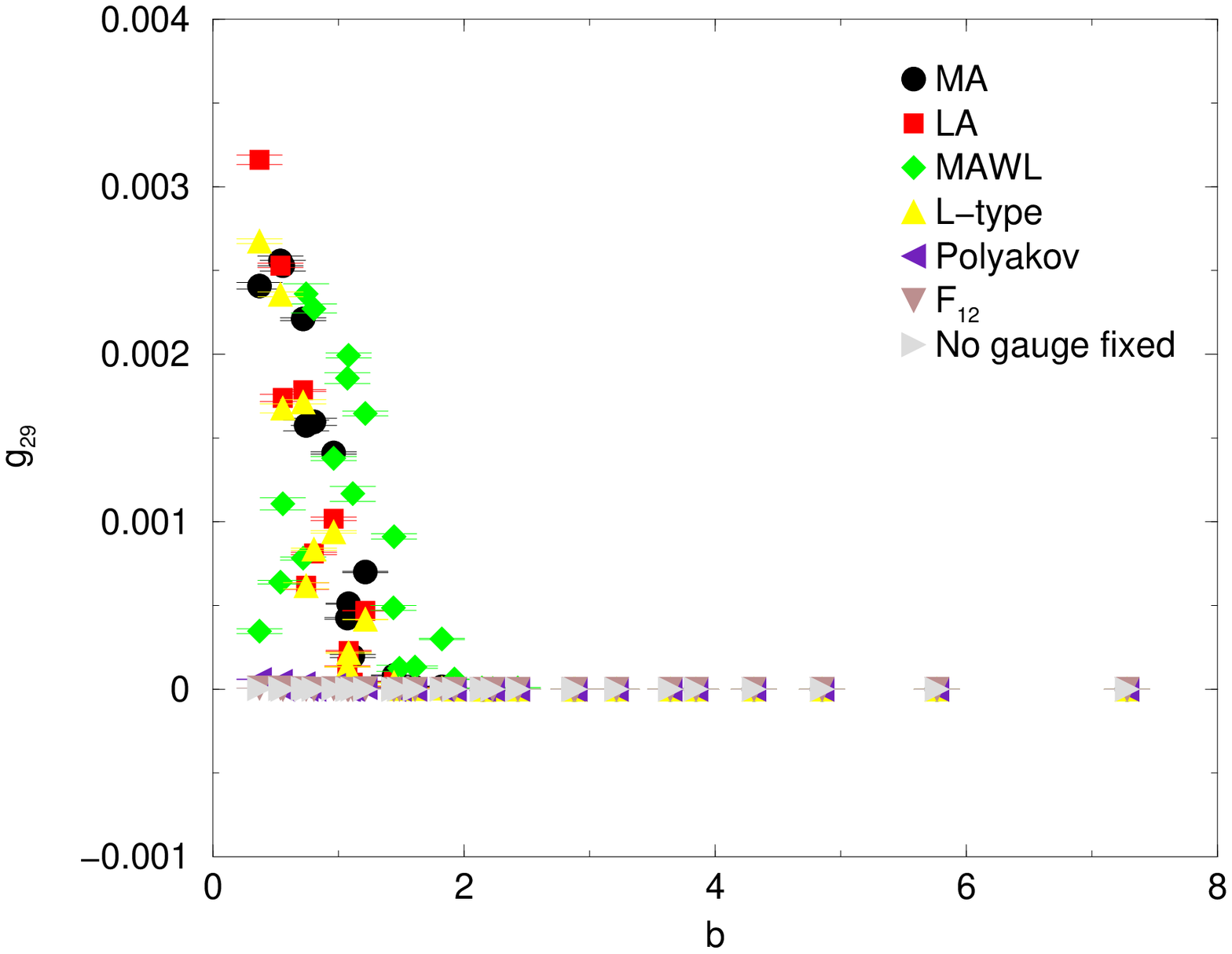,scale=0.41}
{{4-point coupling $g_{28}$ vs. physical scale $b$.}\label{fig-g28_vs_b}}
{{6-point coupling $g_{29}$ vs. physical scale $b$.}\label{fig-g29_vs_b}}

\FIGURE{
\epsfig{scale=0.55,file=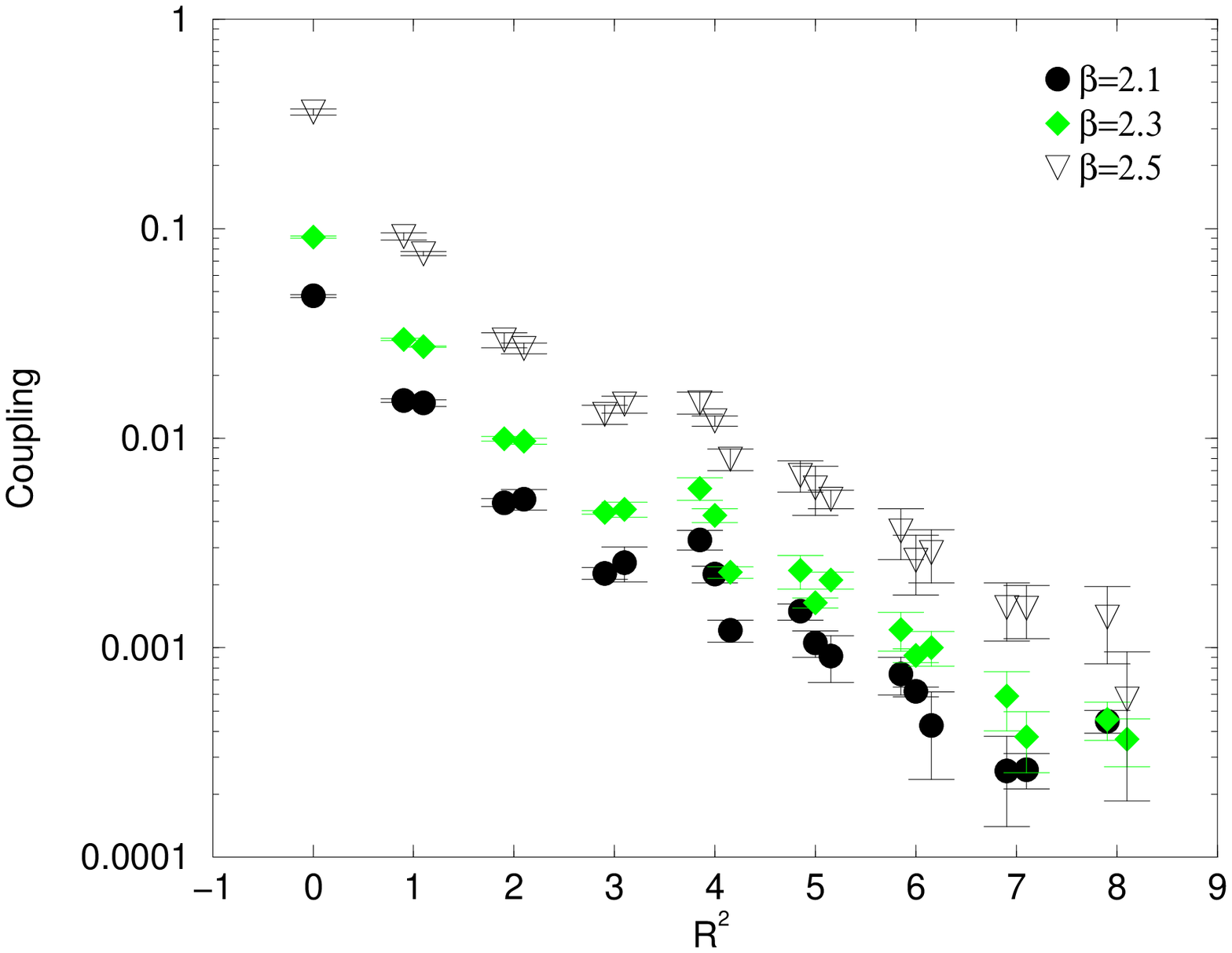}
\caption{Effective couplings vs. squared distances in lattice unit.
(MA gauge, $\beta=2.1,2.3$ and $2.5$, effective couplings for $n=8$ blocked monopole)}
\label{fig-r-dep}
}
}
\item{
We used a standard iterative gauge fixing procedure for MA, MAWL and
L-type gauges.  In such a case, gauge fixing sweeps may be stuck for
some local minima of a gauge fixing functional.  Different local
minima give rise to different gauge transformations, but they can not
be distinguished from the view point of the iterative gauge fixing
procedure.  These are the lattice Gribov copies.  Indeed, Bali et
al. showed that the effect of such copies to the abelian string
tension is not so small~\cite{Bali2}.  To check the effect of copies
to the effective couplings, we generate 100 of $SU(2)$ configurations
on $24^4$ lattice at $\beta=2.5$.  Then, we generate 7 of gauge
equivalent configurations (i.e., copies) via a random gauge
transformation.  Using these gauge copies, we constructed effective
monopole actions and compared their effective couplings.  Figure
\ref{fig-gcopymag} shows $g_1$ in the case of MA
gauge.  $g_1$ for the different
blocking factors are described in different symbols.  We see some
fluctuations in $g_1$ for MA gauge.  This is nothing but the effect of lattice
Gribov copies.  The effect of the copies, however, are negligibly
small.  Therefore, qualitative analyses which are given later will not
be affected.  In principle, LA gauge does not have such a
copy~\cite{Sijs1}.  Indeed, we confirmed that effective couplings for
LA gauge are not affected by Gribov copies (Figure \ref{fig-gcopylag}).

\DOUBLEFIGURE[b]
{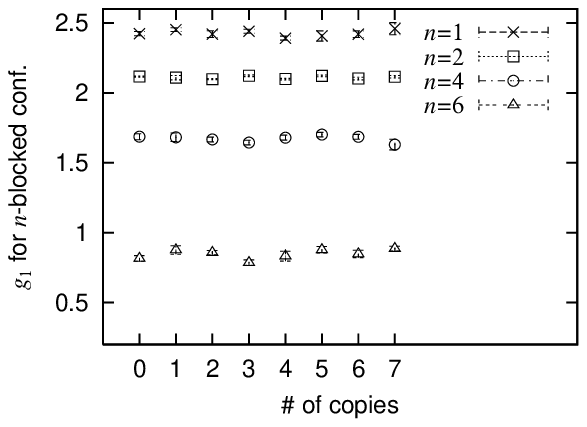,scale=1.2}
{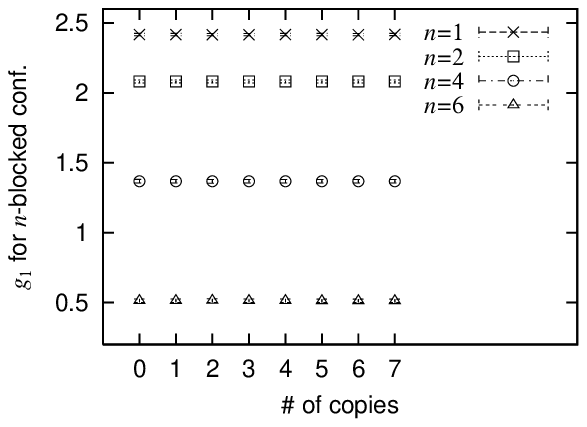,scale=1.2}
{{Gribov copy effect for $g_1$ (MA gauge)}\label{fig-gcopymag}}
{{Gribov copy effect for $g_1$ (LA gauge)}\label{fig-gcopylag}}
}
\item{
Figure \ref{fig-g1vsb1} and Figure \ref{fig-g2vsb1} show the most
dominant quadratic self coupling constant $g_1$ and quadratic
nearest-neighbor coupling constant $g_2$ versus physical scale $b$ in
the case of MA, LA, MAWL and L-type gauges, respectively.  In these
gauges, effective coupling constants take large values in small $b$
region and the scaling behavior (i.e., a unique curve for different
blocking factor $n$) is seen even in small $b$ region.  The effective
actions which are obtained here appear to be a good approximation of
the action on the renormalized trajectory corresponding to the
continuum limit.  In addition to this, coupling constants for these
four gauges are very close to each other, although these gauges have a
completely different form in the continuum limit.
}
\item{
However, in the case of Polyakov, F${}_{12}$ and no gauge fixings,
coupling constants are different from those in the above four gauges
(See, Figure \ref{fig-g1vsb2} and Figure \ref{fig-g2vsb2}).  In these
gauges, coupling constants take smaller values and the scaling
behavior is not seen especially in small $b$ region.  To clarify the
scaling properties of these coupling constants, we show the figures as
showing a distinction between the different blocking factors $n$ in
two typical gauges.  In the case of Polyakov gauge (Figure
\ref{fig-g1vsbpol}), the coupling constants depend on the blocking
factor $n$ strongly in small $b$ regions.  On the other hand, in the
case of LA gauge (Figure \ref{fig-g1vsblag}), renormalized coupling
constants are lying on the unique curve.

\DOUBLEFIGURE[b]
{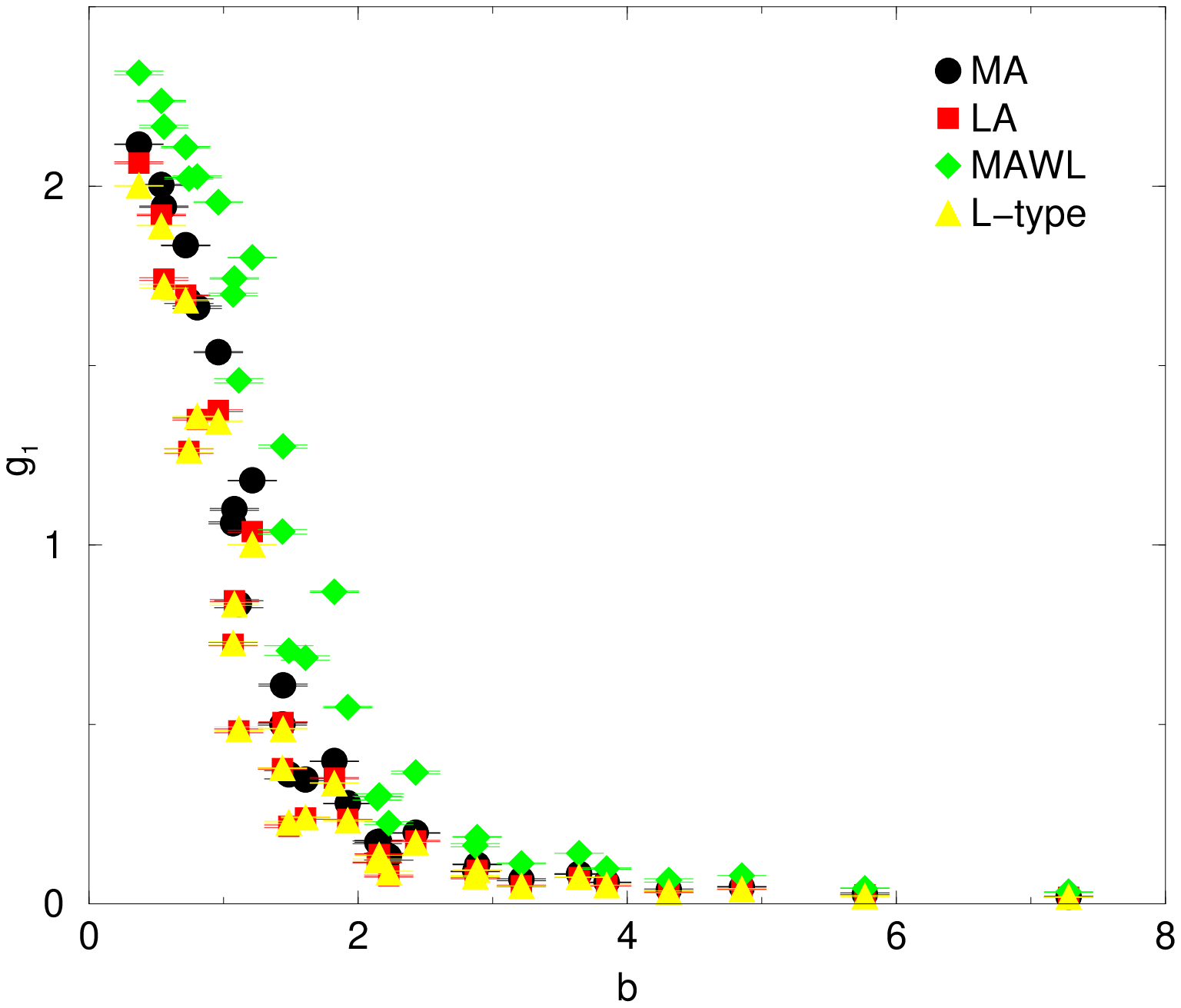,scale=0.41}
{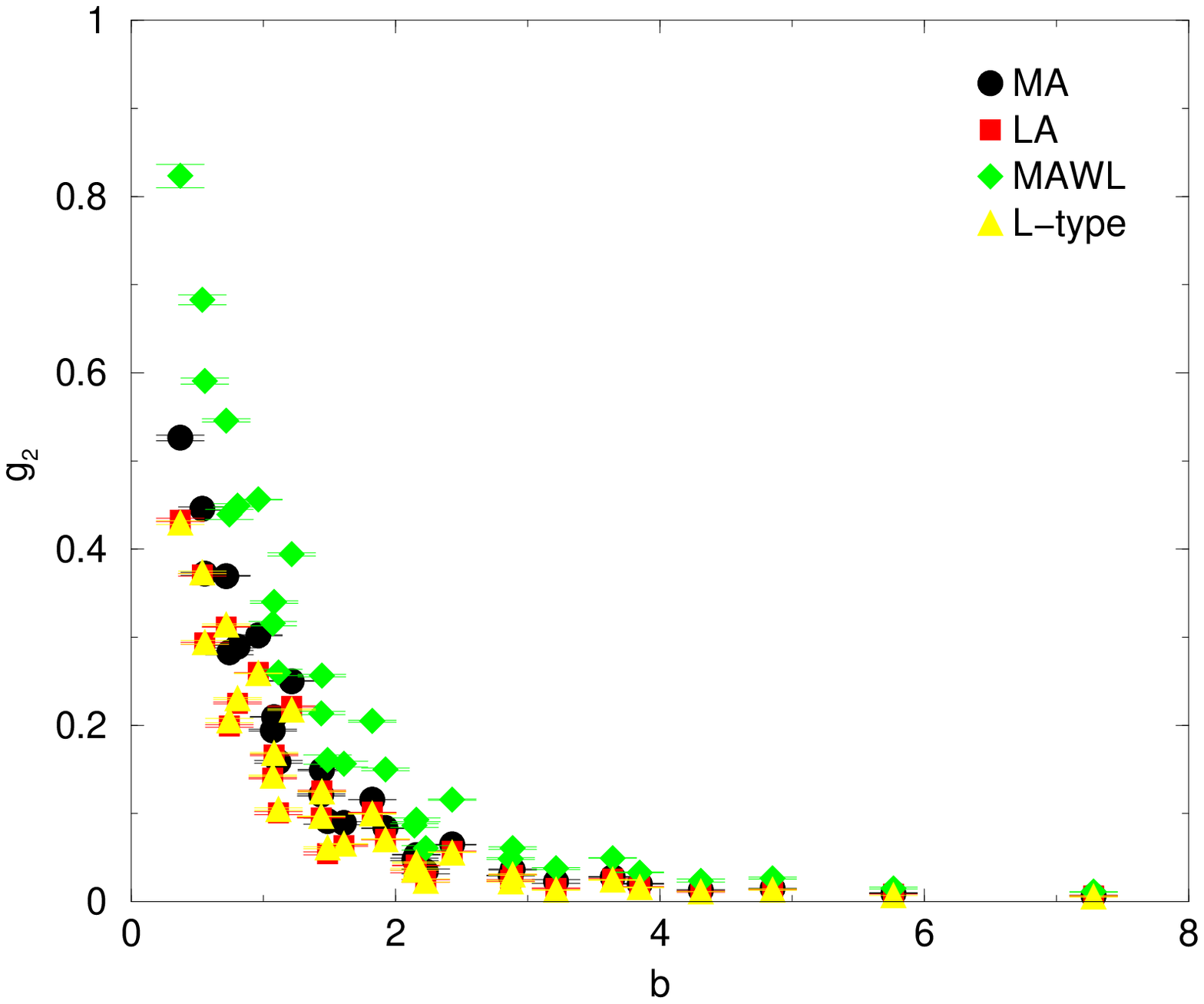,scale=0.41}
{{The most dominant self coupling $g_1$ vs. physical scale $b$ in MA, LA, MAWL and L-type gauges.}\label{fig-g1vsb1}}
{{Nearest-neighbor coupling $g_2$ vs. physical $b$ in MA, LA, MAWL and L-type gauges.}\label{fig-g2vsb1}}

\DOUBLEFIGURE[b]
{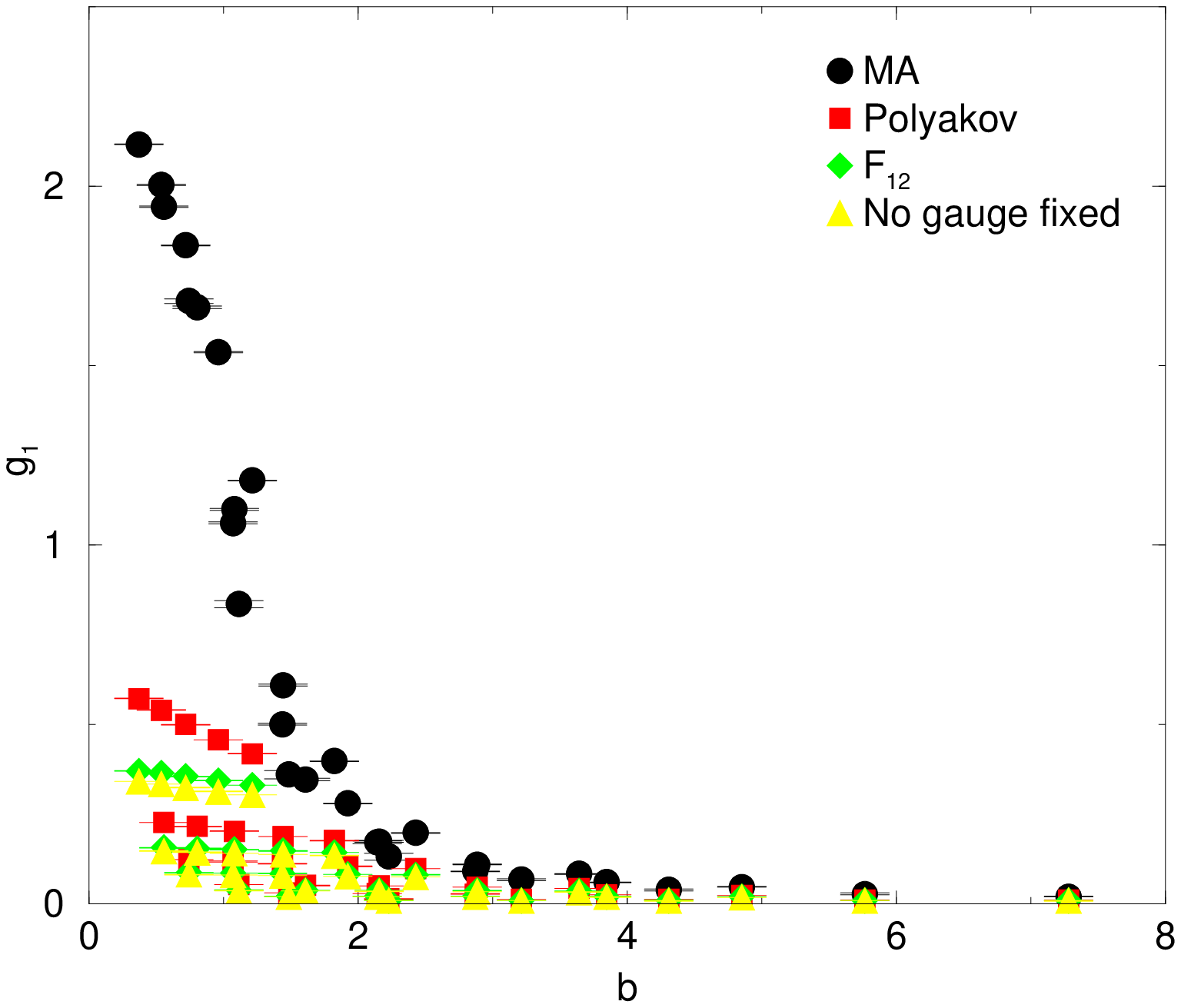,scale=0.41}
{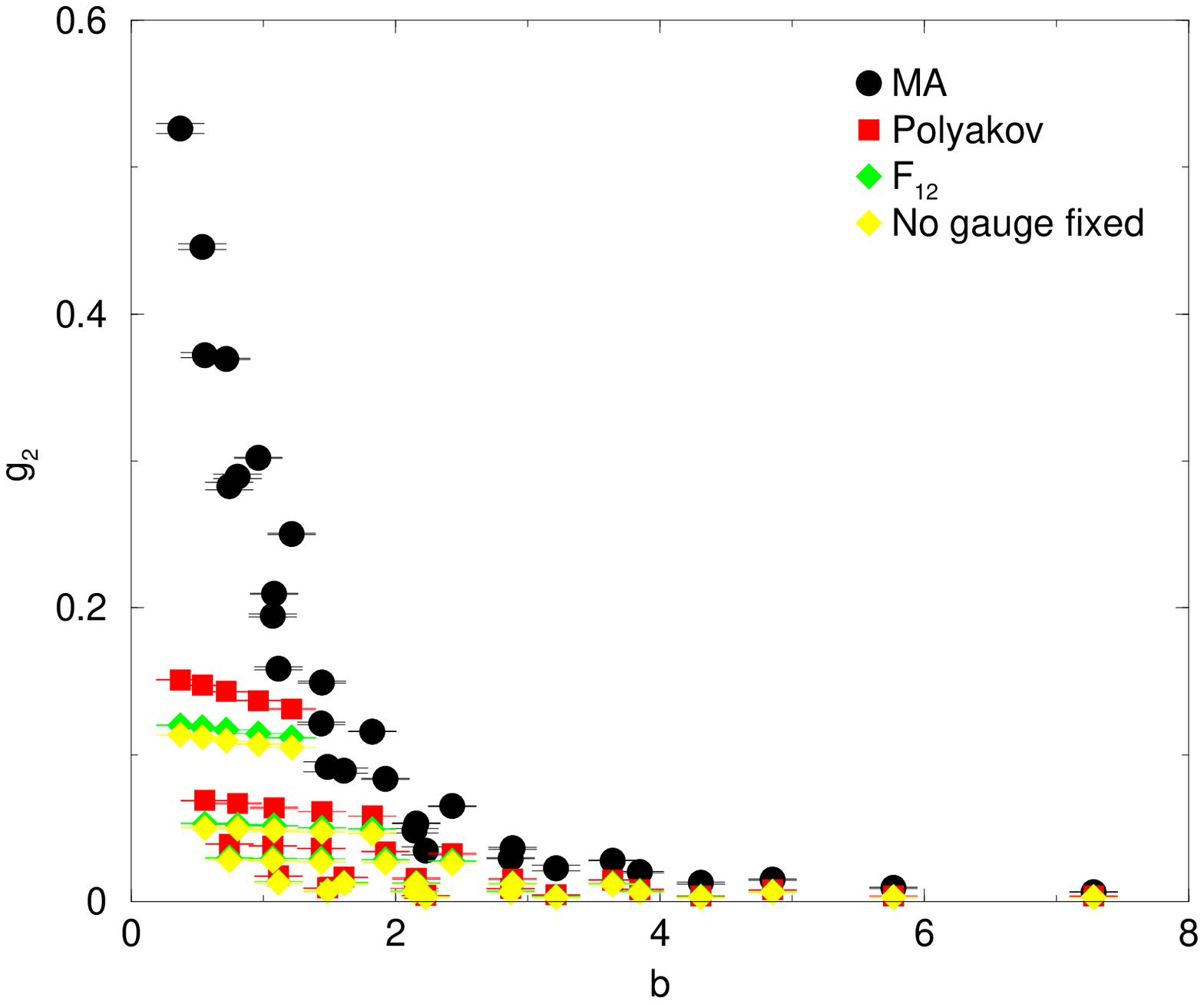,scale=0.41}
{{The most dominant self coupling $g_1$ vs. physical scale $b$ in MA, Polyakov, F${}_{12}$ and no gauge fixings.}\label{fig-g1vsb2}}
{{Nearest-neighbor coupling $g_2$ vs. physical scale $b$ in MA, Polyakov, F${}_{12}$ and no gauge fixings.}\label{fig-g2vsb2}}

\DOUBLEFIGURE[b]
{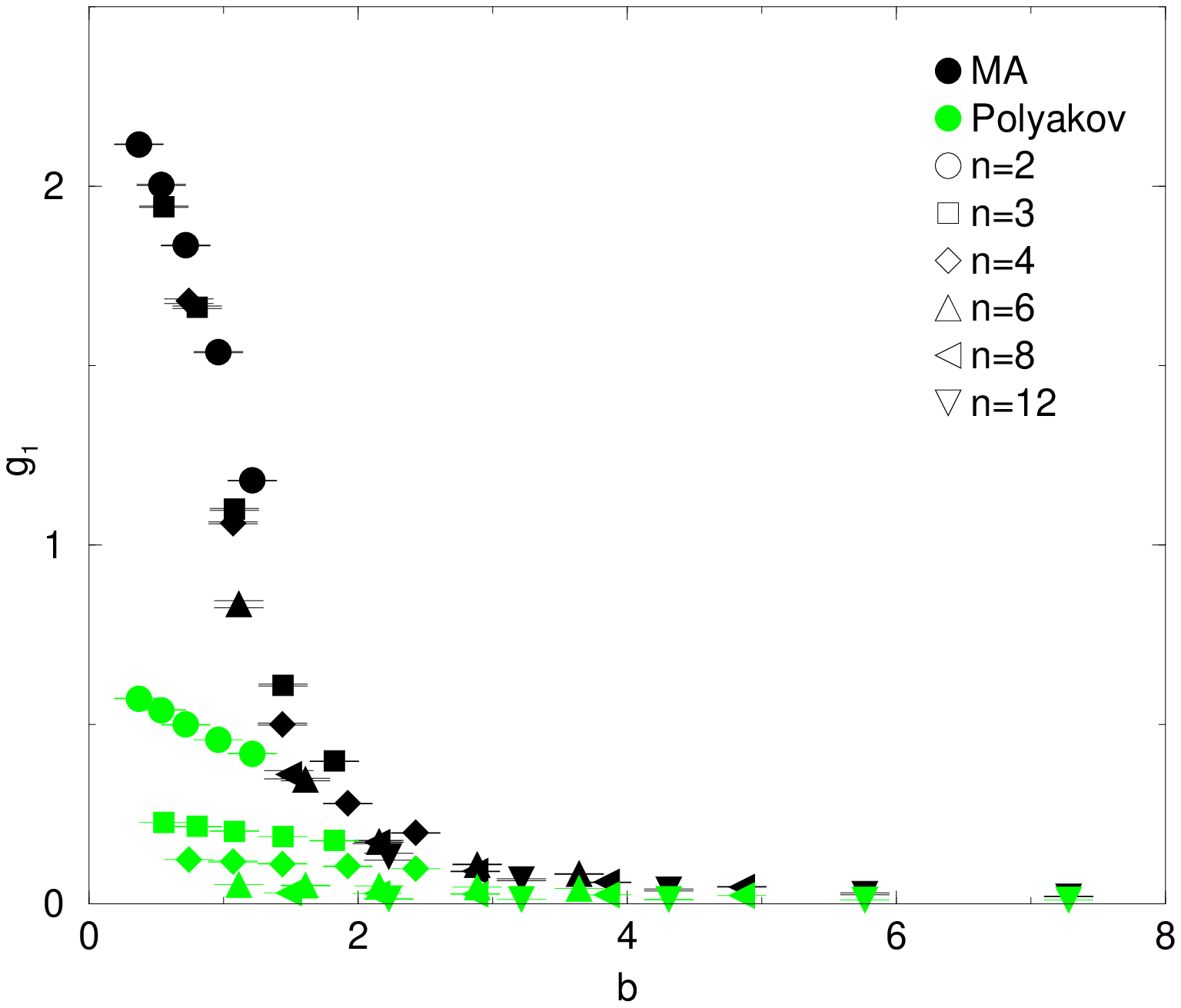,scale=0.41}
{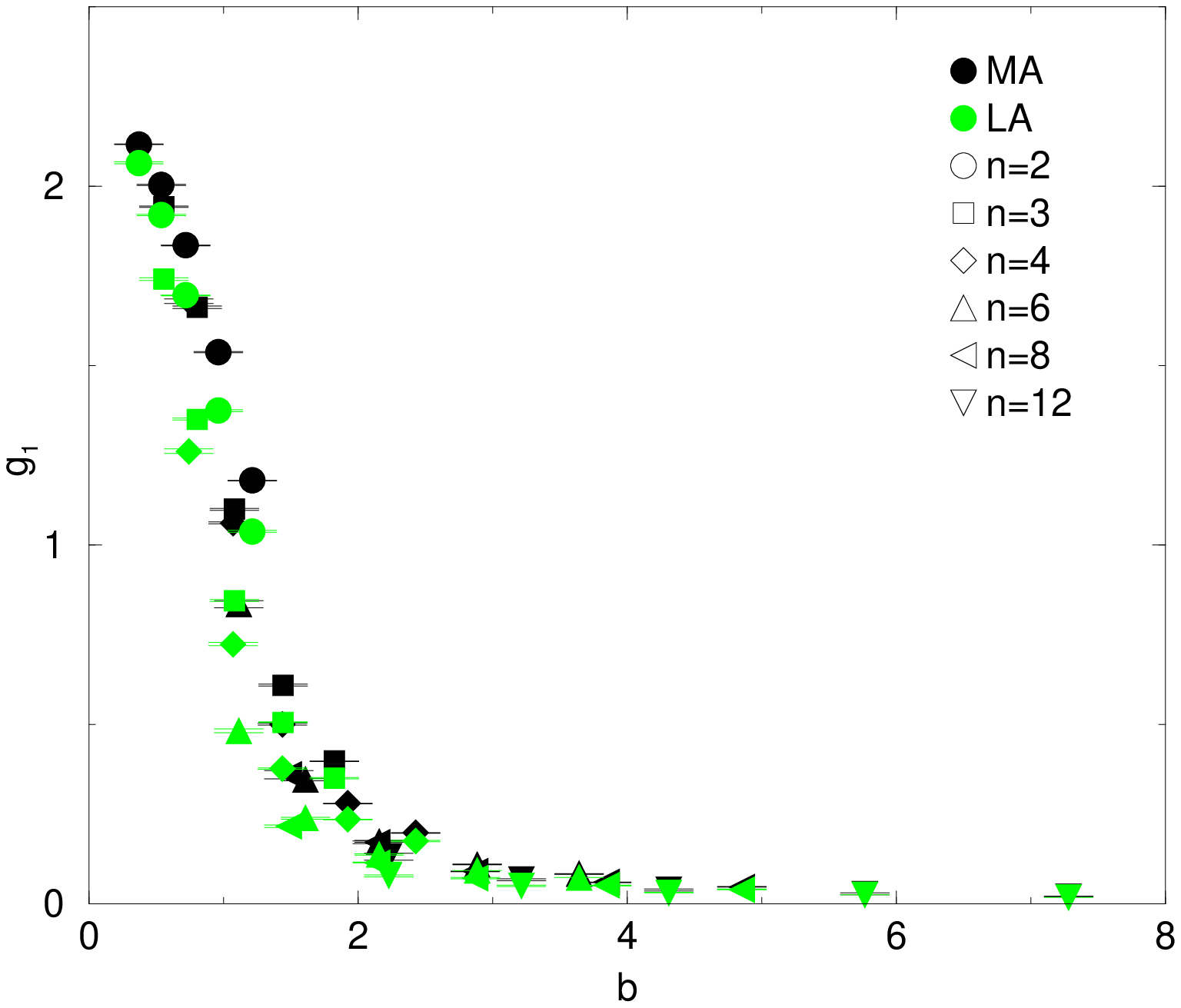,scale=0.41}
{{$g_1$ versus $b$ in MA gauge and Polyakov gauge.  Each symbols correspond to the different blocking factors $n$.}\label{fig-g1vsbpol}}
{{$g_1$ versus $b$ in MA gauge and LA gauge.  Each symbols correspond to the different blocking factors $n$.}\label{fig-g1vsblag}}
}
\item{
Once the effective actions are fixed, we can see from energy-entropy
balance of the system whether monopole condensation occurs or not.  If
the entropy of a monopole loop exceeds the energy, the monopole loop
condenses in the QCD vacuum.  In four-dimensional lattice theory, the
entropy of a monopole loop can be estimated as $\ln 7$ per unit loop
length.  It is determined by the random walk without backward
tracking.  The action can be approximated by the self interaction term
$g_1$ alone since the interactions with two separate currents are
almost canceled~\cite{KitaharaD}.  The free energy per unit monopole
length is approximated by
\[ F\sim g_1-\ln 7, \]
since $g_1$ can be regarded as the self energy per unit monopole loop
length.  If $g_1<\ln 7$, the entropy dominates over the energy, that
is, monopole condensation occurs.  In Figure \ref{fig-g1vsb1}
and Figure \ref{fig-g1vsb2}, we see that the entropy of the system
dominates over the energy in the large $b$ region for all gauges.  In
other words, monopole condensation occurs~\cite{SS1} in the large $b$
region for all gauges.
}
\item{
Figures \ref{fig-flow1},\ref{fig-flow2},\ref{fig-flow3} and
\ref{fig-flow4} show the RG flows projected onto $g_1$-$g_2$,
$g_1$-$g_5$, $g_1$-$g_7$ and $g_1$-$g_{10}$ coupling planes,
respectively.  The effective coupling constants for all gauges seem to
converge to the identical line for large $b$ region.  This may show
gauge independence of the monopole condensation in the low energy
region.  Although all coupling constants become very small in the
large $b$ region, it is important that the slopes of the
renormalization flows seem to converge in all gauges.

\DOUBLEFIGURE[b]
{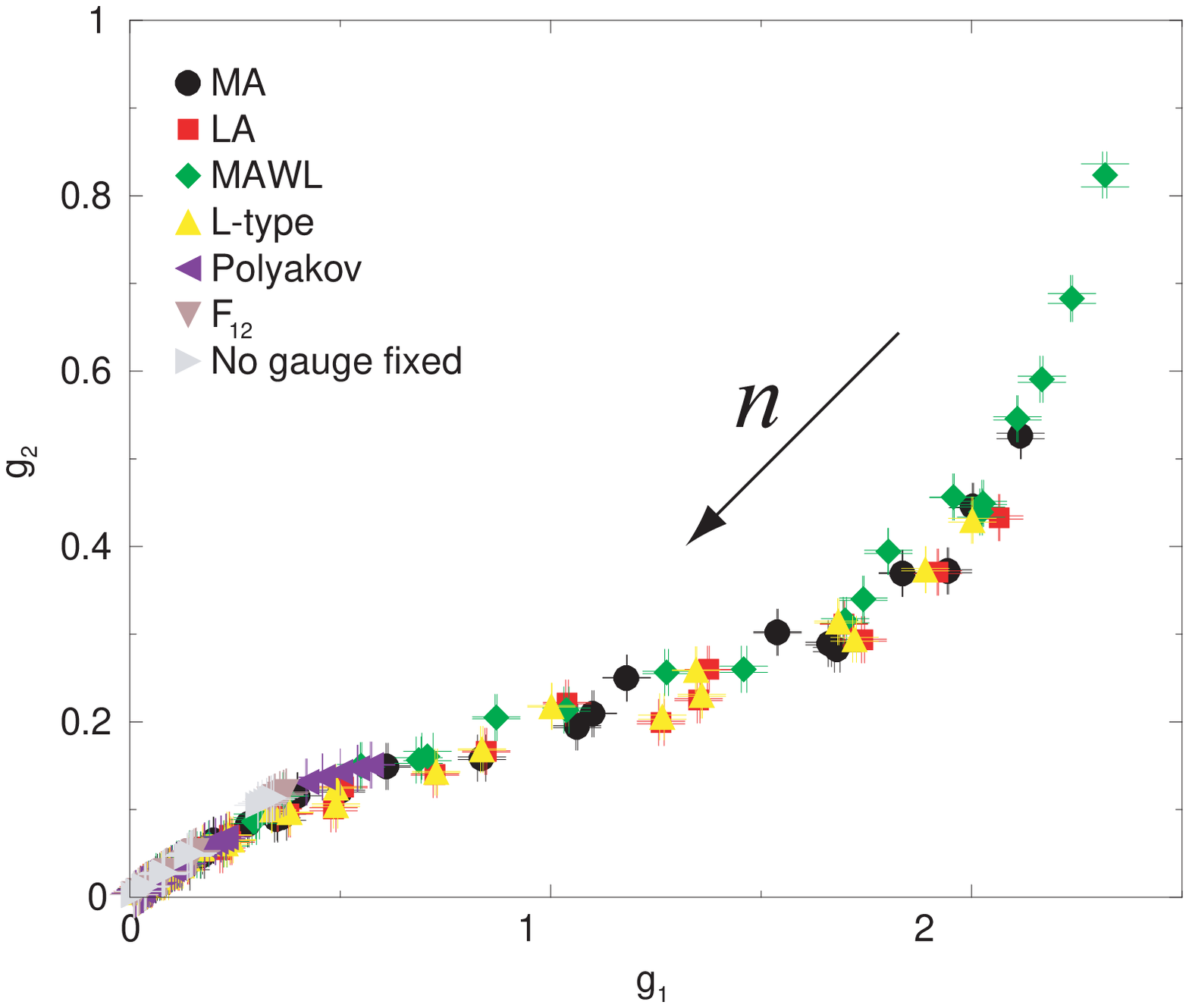,scale=0.41}
{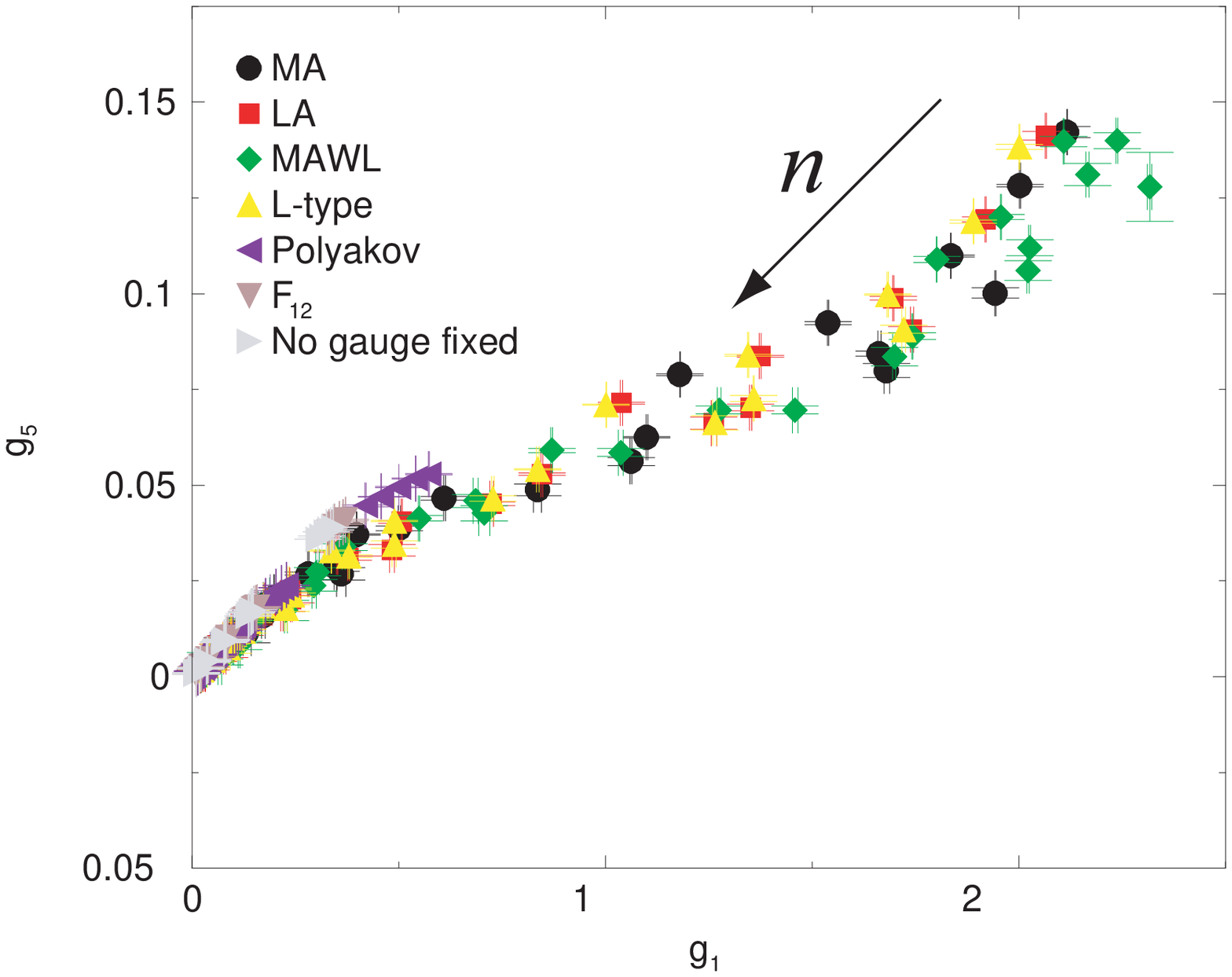,scale=0.41}
{{RG flows of the effective monopole actions project onto the $g_1$-$g_2$ coupling plane}\label{fig-flow1}}
{{RG flows of the effective monopole actions project onto the $g_1$-$g_5$ coupling plane}\label{fig-flow2}}

\DOUBLEFIGURE[b]
{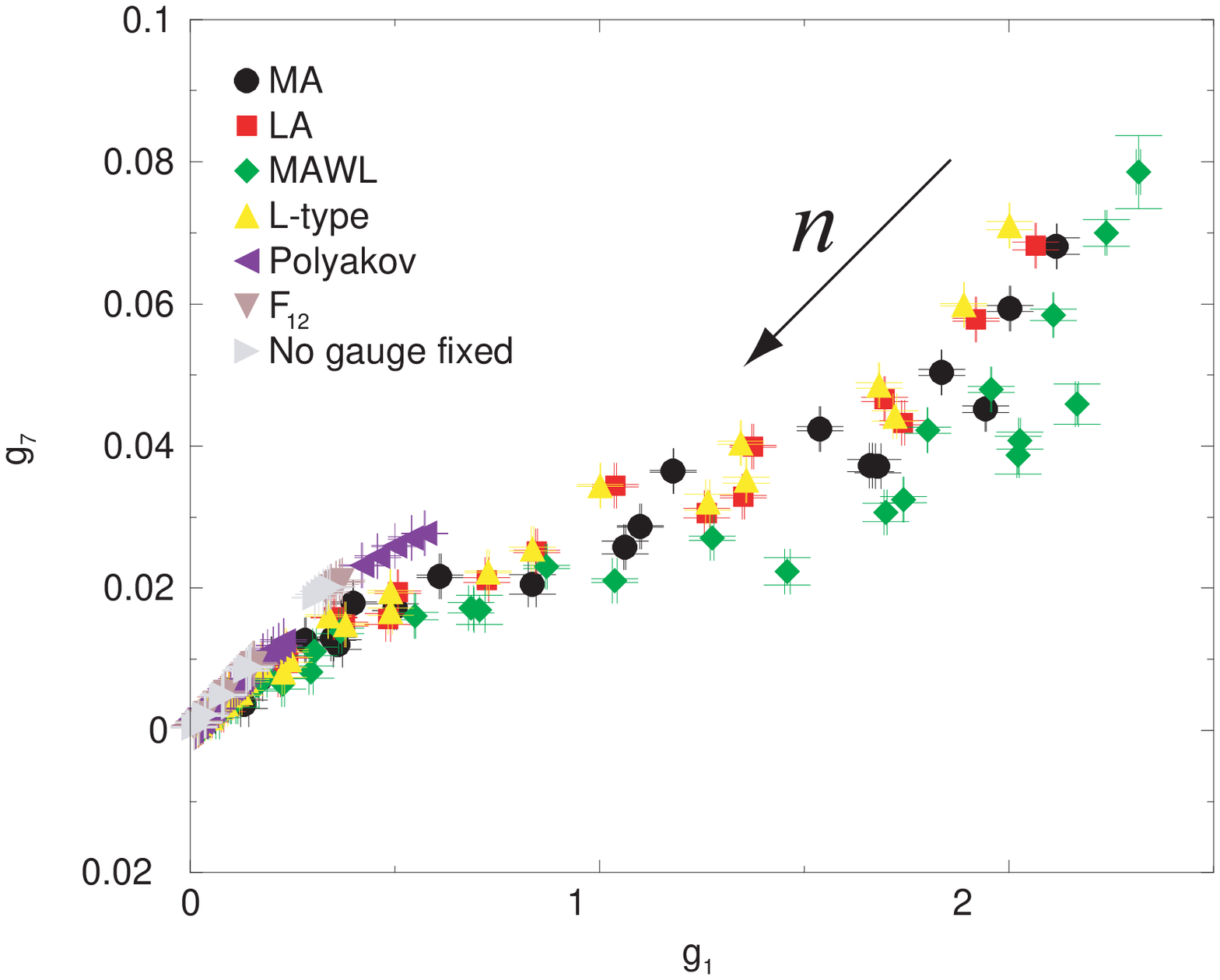,scale=0.41}
{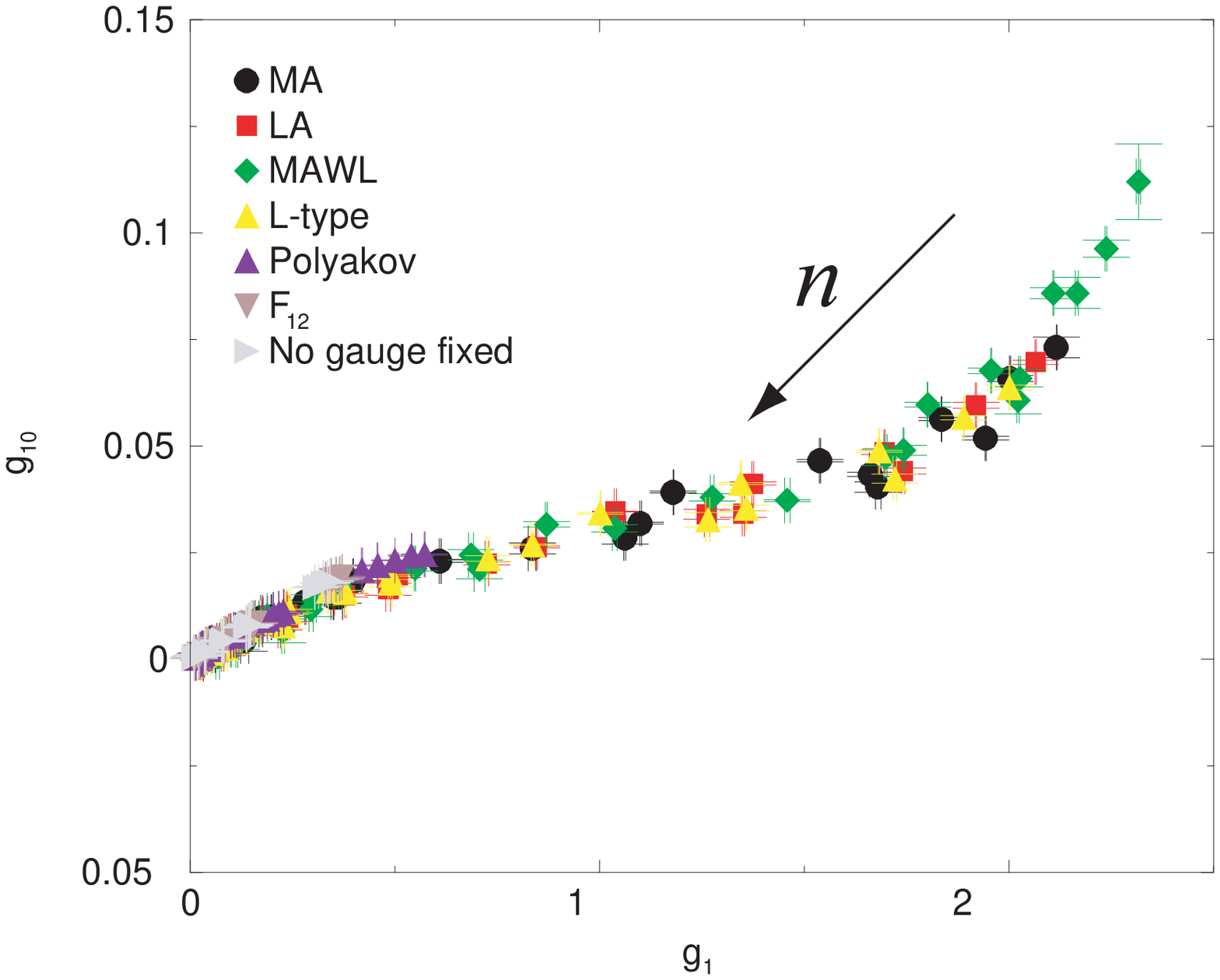,scale=0.41}
{{RG flows of the effective monopole actions project onto the $g_1$-$g_7$ coupling plane}\label{fig-flow3}}
{{RG flows of the effective monopole actions project onto the $g_1$-$g_{10}$ coupling plane}\label{fig-flow4}}
}
\end{enumerate}


\section{Summary}
We have measured first the abelian and the monopole contributions to
the string tension in four types of abelian projections, i.e., MA, LA,
MAWL and L-type gauges.  They show a good agreement with each other.
Monopole string tension are extracted in the same manner as abelian
string tension, and they agree also with each other.  MA and LA gauges
are not unique good gauges.

Next, we have determined the effective monopole actions in various
gauges from monopole vacua using the modified Swendsen's method.  In
the case of MA gauge, an effective monopole action has already been
obtained in Ref.~\cite{SS1}.  In addition to this action, the
effective monopole actions in Polyakov gauge, F${}_{12}$ gauge, LA
gauge, MAWL gauge, L-type gauge and no gauge fixing are also
determined for the first time in this paper.  Moreover, these
effective actions are determined on the blocked monopole vacua, too.
In these effective actions, two point interactions are dominant,
whereas 4-point and 6-point effective coupling constants are
negligibly small in the infrared region.  The RG flows seem to
converge to the identical line when repeating the blockspin
transformation.  It is important that the slopes of renormalization
flows in all gauges seem to converge.  The data are compatible with
the assumption of gauge independence of monopole dynamics in the
continuum limit.  Energy-entropy balance also tells us the monopole
condensation occurs in the large $b$ region for all gauges.


\acknowledgments
The authors would like to thank Fumiyoshi Shoji at Hiroshima
University for fruitful discussions.  This work is supported by the
Supercomputer Project of the Institute of Physical and Chemical
Research (RIKEN).  A part of our numerical simulations have been done
using NEC SX-5 at Research Center for Nuclear Physics (RCNP) of Osaka
University.


\appendix
\section{Maximally Abelian Wilson Loop (MAWL) gauge}
\label{sec-MAWL}
$SU(2)$ gauge field $U_\mu(s)$ can be parameterized by its isospin
components.  In this section, we denote each isospin component of
$U_\mu(s)$ as $U_0(s,\mu)$, $U_1(s,\mu)$ and so on, for simplicity.
This gauge is realized with maximizing the abelian Wilson loop of
$1\times 1$ size:
\begin{equation}
R=\sum_{s,\mu\ne\nu}\cos\Theta_{\mu\nu}(s),
\end{equation}
where the abelian link field is extracted as
\begin{equation}
\theta(s,\mu)=\arctan(U_3(s,\mu)/U_0(s,\mu)).
\end{equation}
Let us consider an infinitesimal gauge transformation of $U$,
\begin{equation}
U'(s,\mu)=(1+i\alpha_i(s)\sigma_i)(U_0(s,\mu)I+iU_j(s,\mu)\sigma_j)
(1-i\alpha_k(s+\hat{\mu})\sigma_k).
\end{equation}
This gives
\begin{eqnarray}
\delta U_0(s,\mu)&=&-(\alpha_i(s)-\alpha_i(s+\hat{\mu}))U_i(s,\mu),\\
\delta U_k(s,\mu)&=&(\alpha_k(s)-\alpha_k(s+\hat{\mu}))U_0(s,\mu)
-\epsilon_{ijk}(\alpha_i(s)+\alpha_i(s+\hat{\mu}))U_j(s,\mu).
\end{eqnarray}
Then $R$ changes as
\begin{equation}
\delta R=-\sum_{s,\mu\ne\nu}\sin\Theta_{\mu\nu}(s)
(\delta\theta(s,\mu)+\delta\theta(s+\hat{\mu},\nu)-
\delta\theta(s+\hat{\mu},\mu)-\delta\theta(s,\nu)),
\end{equation}
where
\begin{equation}
\delta\theta(s,\mu)=\frac{U_0(s,\mu)\delta U_3(s,\mu)-U_3(s,\mu)\delta U_0(s,\mu)}{U_0^2(s,\mu)+U_3^2(s,\mu)}.
\end{equation}

One can check that $R$ is invariant under the $U(1)$ transformation.
Hence we do not need to consider the $\alpha_3(s)$ part.  First, let
us consider the $\alpha_1$ part.  Since there is the sum over whole
lattice sites $s$, one can shift the site variable, for example, $s$
to $s-\hat{\mu}$.  Also one can use the (anti)symmetric property with
respect to $\mu$ and $\nu$ directions.  Finally one gets,
\begin{eqnarray*}
-\frac{\delta R}{2}&=&\sum_{s,\mu\ne\nu}(\alpha_1(s)X_1(s,\mu,\nu)+\alpha_2(s)X_2(s,\mu,\nu)),\\
X_1(s,\mu,\nu)&=&
W_1\frac{U_1(s,\mu)U_3(s,\mu)-U_0(s,\mu)U_2(s,\mu)}{U_0^2(s,\mu)+U_3^2(s,\mu)}\\
&&+W_2\frac{U_1(s-\hat{\mu},\mu)U_3(s-\hat{\mu},\mu)+U_0(s-\hat{\mu},\mu)U_2(s-\hat{\mu},\mu)}{U_0^2(s-\hat{\mu},\mu)+U_3^2(s-\hat{\mu},\mu)},\\
X_2(s,\mu,\nu)&=&
W_1\frac{U_2(s,\mu)U_3(s,\mu)+U_0(s,\mu)U_1(s,\mu)}{U_0^2(s,\mu)+U_3^2(s,\mu)}\\
&&+W_2\frac{U_2(s-\hat{\mu},\mu)U_3(s-\hat{\mu},\mu)-U_0(s-\hat{\mu},\mu)U_1(s-\hat{\mu},\mu)}{U_0^2(s-\hat{\mu},\mu)+U_3^2(s-\hat{\mu},\mu)},\\
W_1&=&\sin\Theta_{\mu\nu}(s)-\sin\Theta_{\mu\nu}(s-\hat{\nu}),\\
W_2&=&\sin\Theta_{\mu\nu}(s-\hat{\mu}-\hat{\nu})-\sin\Theta_{\mu\nu}(s-\hat{\mu}).
\end{eqnarray*}
When we write $X^\pm=X_1\pm iX_2$, it is easy to see
$X^\pm$ transforms covariantly under the residual $U(1)$.

Finally, one gets the matrix which is diagonalized in this gauge,
\begin{eqnarray*}
X(s)&=&\sum_{\mu\ne\nu}\left[
\frac{\sin\Theta_{\mu\nu}(s)-\sin\Theta_{\mu\nu}(s-\hat{\nu})}{U_0^2(s,\mu)+U_3^2(s,\mu)}(U(s,\mu)\sigma_3U^\dagger(s,\mu))\right.\\
&&+\left.\strut\frac{\sin\Theta_{\mu\nu}(s-\hat{\mu}-\hat{\nu})-\sin\Theta_{\mu\nu}(s-\hat{\mu})}{U_0^2(s-\hat{\mu},\mu)+U_3^2(s-\hat{\mu},\mu)}(U^\dagger(s-\hat{\mu},\mu)\sigma_3U(s-\hat{\mu},\mu))
\right].
\end{eqnarray*}
Because of the non-locality of the gauge condition, one can not
calculate the gauge transformation matrix which diagonalizes $X(s)$ in
a simple way.  Therefore, we employed an iterative updation procedure
to satisfy the gauge condition.
\begin{enumerate}
\item{
Make a trial gauge transformation, adopting $\alpha_1$ and
$\alpha_2$ as follows:\\
$\alpha_1(s)=-\kappa X_1(s)$,\\
$\alpha_2(s)=-\kappa X_2(s)$.
}
\item{
Measure $R$.  If $R$ becomes larger than before,
accept this trial and repeat step 1.  If $R$ becomes
smaller than before, take $\kappa_{new}=\kappa_{old}/2$ and adopt the
gauge transformation using this $\kappa_{new}$ with
respect to the configuration
before trial, and then repeat step 1.
}
\item{
If the off-diagonal element of $X(s)$ becomes smaller than a suitable
threshold (we set this to $1.0$), one can regard the gauge fixing
procedure as having been completed.
}
\end{enumerate}
We set an initial value of $\kappa$ to $0.1$.  $R$ can be maximized as
long as we take $\kappa>0$.  We apply the MA gauge fixing as a
preconditioning for MAWL gauge fixing and then we perform the above
procedure to the MA fixed configuration.  This preconditioning is
required to improve a convergence property of the MAWL gauge fixing.
We have to note that the configurations which obtained via the above
procedure are not perfectly gauge fixed because the off-diagonal
element of $X(s)$ still remain not very small.

\end{document}